\documentclass[12pt]{article}
\pdfoutput=1
\usepackage{graphicx}
\DeclareGraphicsExtensions{.pdf}
\usepackage{float} 
\usepackage{amsmath}
\usepackage{amsfonts}   
\usepackage{amssymb}
\usepackage{subfloat}

\begin{document}
\date{}
\title{{\bf{\Large Critical phenomena in higher curvature charged AdS black holes}}}

\author{
{\bf {\normalsize Arindam Lala}$
$\thanks{arindam.lala@bose.res.in}}\\
{\normalsize S.N. Bose National Centre for Basic Sciences,}\\
{\normalsize JD Block, Sector III, Salt Lake, Kolkata 700098, India}\\[0.3cm]
\\[0.3cm]
}
\date{}

\maketitle

\begin{abstract}
In this paper we have studied the critical phenomena in higher curvature charged black holes in the anti-de Sitter (AdS) space-time. As an example we have considered the third order Lovelock-Born-Infeld black holes in seven dimensional AdS space-time. We have analytically derived the thermodynamic quantities of the system. Our analysis revealed the onset of a higher order phase transition in the black hole leading to an infinite discontinuity in the specific heat at constant charge at the critical points. Our entire analysis is based on the canonical framework where we have fixed the charge of the black hole. In an attempt to study the behaviour of the thermodynamic quantities near the critical points, we have derived the critical exponents of the system explicitly. Although, the values of the critical points have been determined numerically, the critical exponents are calculated analytically. Our results fit well with the thermodynamic scaling laws. The scaling hypothesis is also seen to be consistent with these scaling laws. We find that all types of AdS black holes, studied so far, indeed belong to the same universality class. Moreover, these results are consistent with the mean field theory approximation. We have also derived the suggestive values of the other two critical exponents associated with the correlation function and correlation length on the critical surface. 
\end{abstract}
\section{Introduction}
Constructing gravity theories in higher dimensions (i.e. greater than four) has been an interesting topic of research for the past several decades. One of the reasons for this is that these theories provide a framework for unifying gravity with other interactions. Since string theory is an important candidate for such a unified theory, it is necessary to consider higher dimensional space-time for its consistency. In fact, the study of string theory in higher dimensions is one of the most important and challenging sectors in high energy physics. String theory requires the inclusion of gravity in order to describe some of its fundamental properties. Other theories like brane theory, theory of supergravity are also studied in higher dimensions. All these above mentioned facts underscored the necessity for considering gravity theories in higher dimensions\cite{Zumino}. The effect of string theory on gravity may be understood by considering a low energy effective action that describes the classical gravity\cite{Green}. This effective action must include combinations of the higher curvature terms and are found to be ghost free\cite{Zwiebach}. In an attempt to obtain the most general tensor that satisfies the properties of Einstein's tensor in higher dimensions, Lovelock proposed an effective action that contains higher curvature terms\cite{Lovelock}. The field equations derived from this action consists of only second derivatives of the metric and hence are free of ghosts\cite{Deser}. In fact, these theories are the most general second order gravity theories in higher dimensions. 

Among various higher dimensional theories of gravity, it is the seven dimensional gravity that has earned repeated interests over the past two decades because of the interesting physics associated with them. For example, 
addition of certain topological terms in the usual Einstein-Hilbert action produces new type of  gravity in seven dimensions. For a particular choice of the coupling constants it is observed that these terms exist in seven dimensional gauged supergravity\cite{seven6}. Apart from these aspects of the supergravity theories in $7$-dimensions, there are several other important features associated with gravity theories in these particular dimensions. For example, there exists an octonionic instanton solution to the seven dimensional Yang-Mills theory having an extension to a solitonic solution in low energy heterotic string theory\cite{seven7}. Emergence of $U$-duality in seven dimensions involves some non-trivial phenomena which are interpreted in $M$-theory\cite{seven8}. Also, study of black holes in $7$-dimensions is important in the context of the AdS/CFT correspondence\cite{seven9}-\cite{seven11}.

There has been much progress in the study of various properties of the black holes both in four as well as in higher dimensions (greater than four) over the past few decades. Black holes in higher dimensions possesses interesting properties which may be absent in four dimensions\cite{Myres}-\cite{DR2}.\footnote{For an excellent review on black holes in higher dimensions see Ref. \cite{Emparan2}.} Among several intriguing properties of black holes we shall be mainly concerned with the thermodynamics of these objects. This motivation primarily arises from the fact that the study of black hole thermodynamics in higher dimensions may provide us informations about the nature of quantum gravity. The limits of validity of the laws of black hole mechanics\cite{Bardeen} may address some interpretations in this regard. Various quantum effects can be imposed in order to check the validity of these laws. For example, if we want to study the effect of back-reaction of quantum field energy on the laws on black hole mechanics, we are required to consider higher curvature interactions\cite{Birrell}. From this point of view it is very much natural to study the effect of higher curvature terms on the thermodynamic properties of black holes. In fact, the higher dimensional Lovelock theory sets a nice platform to study this effect. Moreover, the higher curvature gravity theories in higher dimensions are free from the complications that arise from the higher derivative theories. All these facts motivates us to study the higher curvature gravity theories in higher dimensions.

The study of thermodynamic properties of black holes in anti-de Sitter (AdS) space-time has got renewed attention due to the discovery of phase transition in Schwarzschild-AdS black holes\cite{Hawking-Page}. Since then a wide variety of research have been commenced in order to study the phase transition in black holes\cite{Ban3, DR2}, \cite{Cham1}-\cite{Bibhas}. Now a days, the study of thermodynamics of black holes in AdS space-time is very much important in the context of AdS/CFT duality. The thermodynamics of AdS black holes may provide important informations about the underlying phase structure and thermodynamic properties of CFTs\cite{Witten}. The study of thermodynamic phenomena in black holes requires an analogy between the variables in ordinary thermodynamics and those in black hole mechanics. Recently, R. Banerjee \textit{et.al.} have developed a method\cite{Modak1, Modak2}, based on Ehrenfest's scheme\cite{Zeemansky} of standard thermodynamics, to study the phase transition phenomena in black holes. In this approach one can actually determine the order of phase transition once the relevant thermodynamic variables are identified for the black holes. This method has been successfully applied in the four dimensional black holes in AdS space-time\cite{Modak2}-\cite{Lala} as well as in higher dimensional AdS black holes\cite{Ban3}. 

The behavior of the thermodynamic variables near the critical points can be studied by means of a set of indices known as static critical exponents\cite{Hankey, Stanley}. These exponents are to a large extent universal, independent of the spatial dimensionality of the system and obey thermodynamic scaling laws\cite{Hankey, Stanley}. The critical phenomena has been studied extensively in familiar physical systems like the Ising model (two and three dimensions), magnetic systems, elementary particles, hydrodynamic systems etc.. An attempt to study the critical phenomena in black holes was commenced in the last twenty years\cite{Lousto1}-\cite{Wu2}. Despite all these attempts, a systematic study of critical phenomena in black holes was still lacking. This problem has been circumvented very recently\cite{DR2, DR1}. In these works the critical phenomena have been studied in (3+1) as well as higher dimensional AdS black holes. Also, the critical exponents of the black holes have been determined by explicit analytic calculations.

All the research mentioned above was confined to Einstein gravity. It would then be interesting to study gravity theories in which action involves higher curvature terms. Among the higher curvature black holes, Gauss-Bonnet and Lovelock black holes will be suitable candidates to study. The thermodynamic properties and phase transitions have been studied in the Gauss-Bonnet AdS (GB-AdS) black holes\cite{Cvetic}-\cite{Anninos}. Also, the critical phenomena in the GB-AdS black holes was studied\cite{Cai6}. On the other hand, Lovelock gravity coupled to the Maxwell field was investigated in  \cite{Deh1, Deh2}. The thermodynamic properties of the third order Lovelock-Born-Infeld-AdS (LBI-AdS) black holes in seven space-time dimensions were studied in \cite{Deh1}, \cite{Aiello, Deh3}. But the study of phase transition and critical phenomena have not yet been done in these black holes.

 In this paper we have studied the critical phenomena in the third order LBI-AdS black holes in seven dimensions. We have also given a qualitative discussion about the possibility of higher order phase transition in this type of black holes. We have determined the static critical exponents for these black holes and showed that these exponents obey the static scaling laws. We have also checked the static scaling hypothesis and calculated the scaling parameters. We observe that these critical exponents take the mean field values. From our study of the critical phenomena we may infer that the third order LBI-AdS black holes yield results consistent with the mean field theory approximation. Despite of having some distinct features in the higher curvature gravity theories, for example, the usual area law valid in Einstein gravity does not hold in these gravity theories, the critical exponents are found to be identical with those in the usual Einstein gravity. This result shows that the AdS black holes, studied so far, belong to the same universality class. We have also determined the critical exponents associated with the correlation function and correlation length. However, the values of these exponents are more suggestive than definitive. As a final remark, we have given a qualitative argument for the determination of these last two exponents.

The organization of the paper is as follows: In section 2 we discuss about the thermodynamical variables of the seven dimensional third order LBI-AdS black holes. In section 3 we analyze the phase transition and stability of these black holes. The critical exponents, scaling laws and scaling hypothesis are discussed in section 4. Finally, we have drawn our conclusions in section 5.
\section{Thermodynamic variables of higher curvature charged AdS black holes}
The effective action in the Lovelock gravity in $(n+1)$ dimensions can be written as\footnote{Here we have taken the gravitational constant $G=1$.}\cite{Lovelock},
\begin{equation}
\mathcal{I}=\frac{1}{16\pi}\int d^{n+1}x \sqrt{-g}\sum_{i=0}^{[\frac{n+1}{2}]}\alpha_{i}\mathcal{L}_{i}
\label{eq2.1}
\end{equation}
where $ \alpha_{i} $ is an arbitrary constant and $ \mathcal{L}_{i} $ is the \textit{Euler density} of a $ 2i $ dimensional manifold. In $ (n+1) $ dimensions all terms for which $ i>[(n+1)/2] $ are equal to zero, the term $ i=(n+1)/2 $ is a topological term and terms for which $ i<[(n+1)/2] $ contribute to the field equations. Since we are studying  third order Lovelock gravity in the presence of Born-Infeld nonlinear electrodynamics\cite{Born}, the effective action of eq.\eqref{eq2.1} may be written as,
\begin{eqnarray}
\mathcal{I}&=&\frac{1}{16\pi}\int d^{n+1}x \sqrt{-g}\,\Big(\alpha_{0}\mathcal{L}_{0} +\alpha_{1}\mathcal{L}_{1}+\alpha_{2}\mathcal{L}_{2}+\alpha_{3}\mathcal{L}_{3}+L(F) \Big)\nonumber\\
               &=&\frac{1}{16\pi}\int d^{n+1}x \sqrt{-g}\,\Big(-2 \Lambda +\mathcal{R}+\alpha_{2}\mathcal{L}_{2}+\alpha_{3}\mathcal{L}_{3}+L(F) \Big)
 \label{eq2.2}              
\end{eqnarray}
where $ \Lambda $ is the cosmological constant given by $ -n(n-1)/2l^2 $, $ l $ being the AdS length, $ \alpha_{2} $ and $ \alpha_{3} $ are the second and third order Lovelock coefficients, $\mathcal{L}_{1}= \mathcal{R} $ is the usual Einstein-Hilbert Lagrangian, $ \mathcal{L}_{2}=R_{\mu\nu\gamma\delta}R^{\mu\nu\gamma\delta}-4R_{\mu\nu}R^{\mu\nu}+\mathcal{R}^{2}$ is the Gauss-Bonnet Lagrangian and\\
\begin{eqnarray*}
\mathcal{L}_{3}&=&2R^{\mu\nu\sigma\kappa}R_{\sigma\kappa\rho\tau}
R^{\rho\tau}\,_{\mu\nu}+8R^{\mu\nu}\,_{\sigma\rho}
R^{\sigma\kappa}\,_{\nu\tau}
R^{\rho\tau}\,_{\mu\kappa}+24                      	R^{\mu\nu\sigma\kappa}R_{\sigma\kappa\nu\rho}
R^{\rho}\,_{\mu}+\\
&&3\mathcal{R}R^{\mu\nu\sigma\kappa}
R_{\sigma\kappa\mu\nu}+24R^{\mu\nu\sigma\kappa}R_{\sigma\mu}
R_{\kappa\nu}+16R^{\mu\nu}R_{\nu\sigma}R^{\sigma}\,_{\mu}-
12\mathcal{R}R^{\mu\nu}R_{\mu\nu}+\mathcal{R}^{3}
\end{eqnarray*}
is the third order Lovelock Lagrangian, $ L(F) $ is the Born-Infeld Lagrangian given by,
\begin{equation}
L(F)=4b^{2}\left(1-\sqrt{1+\frac{F^2}{2b^2}} \right).
\label{eq2.3} 
\end{equation}
In eq.\eqref{eq2.3},  $ F_{\mu\nu}=\partial_{\mu}A_{\nu}-\partial_{\nu}A_{\mu} $, $ F^{2}=F_{\mu\nu}F^{\mu\nu} $ and $ b $ is the Born-Infeld parameter. In the limit $ b\rightarrow\infty $ we recover the standard Maxwell form $ L(F)=-F^{2} $.
  
  The solution of the third order Lovelock-Born-Infeld anti de-Sitter black hole (LBI-AdS) in $ (n+1) $-dimensions  can be written as\cite{Deh3},
\begin{equation}
ds^2=-f(r)dt^{2}+\frac{1}{f(r)}dr^{2}+r^{2}d\Omega_{k,n-1}^{2}
\label{eq2.4}
\end{equation}
where $
 d\Omega_{k,n-1}^{2}=
\left\{
\begin{array}{lr}
d\theta_{1}^{2}+\sum_{i=2}^{n-1}\prod_{j=1}^{i-1}sin^{2}
\theta_{j}d\theta_{i}^{2} &k=1\\
d\theta_{1}^{2}+sinh^{2}\theta_{1}d\theta_{2}^{2}+sinh^{2}
\theta_{1}\sum_{i=3}^{n-1}\prod_{j=2}^{i-1}
                               sin^{2}\theta_{j}d\theta_{i}^{2}&k=-1\\
\sum_{i=1}^{n-1}d\phi_{i}^{2}&k=0                               
\end{array}
\right.
$\\
and $ k $ determines the structure of the black hole horizon\footnote{$ k=+1(spherical),-1(hyperbolic),0(planar) $}. At this point of discussion it must be mentioned that the Lagrangian of eq.\eqref{eq2.2} is the most general Lagrangian in seven space-time dimensions that produces the second order field equations\cite{Deh1}. Thus, We shall restrict ourselves in the seven space-time dimensions.

The equation of motion for the electromagnetic field for this ($6+1$)-dimensional space-time can be obtained by varying the action (eq.\eqref{eq2.2}) with respect to the gauge field $A_{\mu}$. This results the following\cite{Deh3}:

\begin{equation}
\partial_{\mu}\left(\dfrac{\sqrt{-g}F^{\mu\nu}}{\sqrt{1+\frac{F^{2}}{2b^{2}}}}\right)=0,
\label{eq2.5}
\end{equation}
which has a solution\cite{Deh3}
\begin{equation}
A_{\mu}=-\sqrt{\frac{5}{8}}\left(\dfrac{q}{r^{4}}\right) \mathcal{H}(\eta)\delta^{0}_{\mu}.
\label{eq2.6}
\end{equation}
Here $\mathcal{H}(\eta)$ is the abbreviation of the hypergeometric function\cite{Handbook} given by,
\begin{equation}
\mathcal{H}(\eta)=\mathcal{H}\left(\frac{1}{2},\frac{2}{5},\frac{7}{5},-\eta \right).
\label{eqH_2.7}
\end{equation}
In eq.\eqref{eq2.6}, $\eta =\frac{10q^{2}}{b^{2}r^{10}}$ and $q$ is a constant of integration which is related to the charge ($Q$) of the black hole. 
The charge ($Q$) of the black hole may be obtained by calculating the flux of the electric field at infinity\cite{Deh1,Deh3, Carroll}. Therefore, 
\begin{eqnarray}
Q&=&-\int_{\mathcal{B}}d^{n-1}\omega\sqrt{\sigma}n_{\mu}\tau_{\nu}\left(\dfrac{F^{\mu\nu}}{\sqrt{1+\frac{F^{2}}{2b^{2}}}}\right)\nonumber\\
&=&\frac{\mathcal{V}_{n-1}}{4\pi}\sqrt{\frac{(n-1)(n-2)}{2}}q\nonumber\\
&=&\dfrac{\sqrt{10}\pi^{2}q}{4}\qquad(\text{for}\;\;n=6).
\label{eq2.8}
\end{eqnarray}

where $n_{\mu}$ and $\tau_{\nu}$ are the time-like and space-like unit normal vectors to the boundary $\mathcal{B}$, respectively; $\sigma$ is the determinant of the induced metric $\sigma_{ij}$ on $\mathcal{B}$ having coordinates $\omega^{i}$. It is to be noted that, in deriving eq.\eqref{eq2.8} we have only considered the $F^{0r}$ component of $F^{\mu\nu}$.

The quantity $\mathcal{V}_{n-1}$ is the volume of the ($n-1$) sphere and may be written as,
\begin{equation}
 \mathcal{V}_{n-1}=\frac{2\pi^{n/2}}{\Gamma(n/2)}.
 \label{eq2.10}
\end{equation}

The metric function $ f(r) $ of eq.\eqref{eq2.4} may be written as\cite{Deh3},
\begin{equation}
f(r)=k+\frac{r^{2}}{\alpha'}\left( 1-\chi (r)^{\frac{1}{3}}\right)
\label{eq2.11}
\end{equation}
where 
\begin{equation}
\chi (r)=1+\frac{3\alpha' m}{r^{6}}-\frac{2\alpha' b^{2}}{5}\left[ 1-\sqrt{1+\eta}-\frac{\Lambda}{2b^2}+\frac{5\eta}{4}\mathcal{H}(\eta)\right] .
\label{eq2.12}
\end{equation}

Here we have considered the special case $\alpha_{3}=2\alpha_{2}^{2}=\frac{\alpha'^{2}}{72} $ \cite{Deh1, Deh3}. 

Since, in our study of the critical phenomena in the third order LBI-AdS black holes, the thermodynamic quantities like `quasilocal energy' ($M$), Hawking temperature ($T$), entropy ($S$) etc. will play important roles, we will now focus mainly on the derivations of these quantities.

The `quasilocal energy' $M$ of asymptotically AdS black holes may be obtained by using the counterterm method which is indeed inspired by the AdS/CFT correspondence\cite{Malda1}-\cite{Malda2}. This is a well known technique which removes the divergences in the action and conserved quantities of the associated space-time. These divergences appear when one tries to add surface terms to the action in order to make it well-defined. The counterterm method was applied earlier for the computation of the conserved quantities associated with space-time having finite boundaries, de-Sitter (dS) space-time and asymptotically flat space-time in the framework of Einstein gravity\cite{Count1}-\cite{Count7}. On the other hand, this method was applied in Lovelock gravity to compute the associated conserved quantities\cite{Deh1, Deh2, Deh3, Deh4, Deh5}. However, for the asymptotically AdS solutions of the third order Lovelock black holes the action may be written as\cite{Deh2, Deh3, Deh5},
\begin{equation}
\mathcal{A}=\mathcal{I}+\underbrace{\frac{1}{8\pi}\int_{\partial\mathcal{M}} d^{n}x\sqrt{|\gamma|}\,\Big(\mathcal{I}_{b}^{1}+\alpha_{2}\mathcal{I}_{b}^{2}+\alpha_{3}
\mathcal{I}_{b}^{3}\Big)}_\text{boundary terms}+\underbrace{\frac{1}{8\pi}\int_{\partial\mathcal{M}}  d^{n}x\sqrt{|\gamma|}\Big(\dfrac{n-1}{L}\Big)}_\text{counterterm},
\label{eqM.1}
\end{equation}

where $\gamma$ is the determinant of the induced metric $\gamma_{ab}$ on the time-like boundary $\partial\mathcal{M}$ of the space-time manifold $\mathcal{M}$. The quantity $L$ is a scale length factor given by
\begin{equation}
L=\dfrac{15\sqrt{\alpha'(1-\lambda)}}{5+9\alpha'-\lambda^{2}-4\lambda}
\label{eqM.2}
\end{equation}
where
\begin{equation}
\lambda =\Big(1-3\alpha'\Big)^{\frac{1}{3}}.
\end{equation}
The boundary terms in eq.\eqref{eqM.1} is chosen such that the action possesses well defined variational principle, whereas, the counterterm makes the action and the associated conserved quantities finite. 

The boundary terms appearing in eq.\eqref{eqM.1} may be identified as\cite{Deh2, Deh3, Deh5},
\begin{eqnarray}
\mathcal{I}_{b}^{1}&=&K\\
\mathcal{I}_{b}^{2}&=&2\Big(J-2\bar{G}^{1}_{ab}K^{ab}\Big)\\
\mathcal{I}_{b}^{3}&=&3\Big(P-2\bar{G}^{2}_{ab}K^{ab}-12\bar{R}_{ab}J^{ab}+2\bar{\mathcal{R}}J\nonumber\\
&&-4K\bar{R}_{abcd}
K^{ac}K^{bd}-8K\bar{R}_{abcd}K^{ac}K^{b}_{e}K^{ed}\Big),
\label{eqM.3}
\end{eqnarray}

where $K$ is the trace of the extrinsic curvature $K_{ab}$. In eq.(16) $\bar{G}^{1}_{ab}$ is the Einstein tensor for $\gamma_{ab}$ in $n$-dimensions and $J$ is the trace of the following tensor:
\begin{equation}
J_{ab}=\dfrac{1}{3}\Big(2KK_{ac}K^{c}_{b}+K_{cd}K^{cd}K_{ab}-2K_{ac}K^{cd}K_{db}-K^{2}K_{ab}\Big).
\end{equation}
In eq.\eqref{eqM.3} $\bar{G}^{2}_{ab}$ is the second order Lovelock tensor for $\gamma_{ab}$ in $n$-dimensions which is given by
\begin{equation}
\bar{G}^{2}_{ab}=2\Big(\bar{R}_{acde}\bar{R}^{\;\; cde}_{b}-2\bar{R}_{afbc}\bar{R}^{fc}-2\bar{R}_{ac}\bar{R}^{c}_{\;\;b}+\mathcal{\bar{R}}\bar{R}_{ab}\Big)-
\mathcal{L}_{2}(\gamma)\gamma_{ab},
\label{eqM.4}
\end{equation}
whereas, $P$ is the trace of 
\begin{eqnarray}
P_{ab}&=&\dfrac{1}{5}\Bigg( \Big[K^{4}-6K^{2}K^{cd}K_{cd}+8KK_{cd}K^{d}_{e}K^{ec}-6K_{cd}K^{de}K_{ef}
K^{fc}+3\Big(K_{cd}K^{cd}\Big)^{2} \Big]K_{ab}\nonumber\\
&&-\Big(4K^{3}-12KK_{ed}K^{ed}+8K_{de}K^{e}_{f}K^{fd}\Big)K_{ac}K^{c}_{b}-24KK_{ac}K^{cd}K_{de}K^{e}_{b}
\nonumber\\
&&+\Big(12K^{2}-12K_{ef}K^{ef}\Big)K_{ac}K^{cd}K_{db}+24K_{ac}K_{cd}K_{de}K^{ef}K_{fb}\Bigg).
\label{eqM.5}
\end{eqnarray}

Using the method prescribed in Refs.\cite{Brown} we obtain the divergence free energy-momentum tensor as\cite{Deh2, Deh3, Deh5},
\begin{equation}
T^{ab}=\dfrac{1}{8\pi}\Big[\Big(K^{ab}-K\gamma^{ab}\Big)+2\alpha_{2}\Big(3J^{ab}-J\gamma^{ab}\Big)+
3\alpha_{3}\Big(5P^{ab}-P\gamma^{ab}\Big)+\dfrac{n-1}{L}\gamma^{ab}\Big]
\label{eqM.6}
\end{equation}

The first three terms of eq.\eqref{eqM.6} result from the variation of the boundary terms of eq.\eqref{eqM.1} with respect to the induced metric $\gamma^{ab}$, whereas, the last term is obtained by considering the variation of the counterterm of eq.\eqref{eqM.1} with respect to $\gamma^{ab}$. 

For any space-like surface $\mathcal{B}$ in $\partial\mathcal{M}$ which has the metric $\sigma_{ij}$ we can write the boundary metric in the following form\cite{Deh2, Deh3, Deh5, Brown}:
\begin{equation}
\gamma_{ab}dx^{a}dx^{b}=-N^{2}dt^{2}+\sigma_{ij}\Big(d\omega^{i}+V^{i}dt\Big)\Big(d\omega^{j}+V^{j}dt\Big).
\label{eqM.6a}
\end{equation}
where $\omega^{i}$ are coordinates on $\mathcal{B}$, $N$ and $V^{i}$ are the lapse function and the shift vector respectively. 
For any Killing vector field $\xi$ on the the space-like boundary $\mathcal{B}$ in $\partial\mathcal{M}$, we may write the conserved quantities associated with the energy momentum tensor ($T^{ab}$) as\cite{Deh2, Deh3, Deh5, Brown},
\begin{equation}
\mathcal{Q}_{\xi}=\int_{\mathcal{B}}d^{n-1}\omega \sqrt{\sigma}\;n^{a}\xi^{b}T_{ab}
\label{eqM.7}
\end{equation}
where $\sigma$ is the determinant of the metric $\sigma_{ij}$ on $\mathcal{B}$, $n^{a}$ is the time-like unit normal to $\mathcal{B}$ and $\xi^{b}(=\frac{\partial}{\partial t})$ is the time-like Killing vector field. For the metric eq.\eqref{eq2.4} we can write $n^{a}=(1/\sqrt{f(r)},0,0,0,...)$,
$\xi^{a}=(1,0,0,....)$, 
$K_{ab}=-\gamma^{m}_{a}\nabla_{m}\tau_{b}$, where $\tau^{a}=(0,f(r),0,....)$ is the space-like unit normal to the boundary.

With these values of $n^{a}$ and $\xi^{a}$ the only nonvanishing component of $T_{ab}$ becomes $T_{00}$. Hence, $\mathcal{Q}_{\xi}$ corresponds to the `quasilocal energy' $M$ of the black hole. Thus, from eq.\eqref{eqM.7} the expression for the `quasilocal energy' of the black hole may be written as,
\begin{equation}
M=\int_{\mathcal{B}}d^{n-1}\omega \sqrt{\sigma}\;n^{0}\xi^{0}T_{00}.
\label{eqM.8}
\end{equation}

Using eqs.\eqref{eq2.11} and \eqref{eqM.6}, eq.\eqref{eqM.8} can be computed as,
\begin{eqnarray}
M\Big |_{n=6}&=&\dfrac{\mathcal{V}_{n-1}}{16\pi}(n-1)m\Bigg |_{n=6}\nonumber\\
&=&\dfrac{5\pi^{2}}{16}m
\label{eqM.9}
\end{eqnarray} 
where the constant $m$ is expressed as the real root of the equation
\begin{equation}
f(r=r_{+})=0.
\label{eqM.10}
\end{equation}

Using eq.\eqref{eq2.10} and substituting $m$ from eq.\eqref{eqM.10} we finally obtain from eq.\eqref{eqM.9}
\begin{equation}
M=\frac{5\pi^{2}}{16}\Bigg[\frac{k^{3}\alpha'^{2}}{3}+kr_{+}^{4}+k^{2}\alpha' r_{+}^{2}+\frac{2b^{2}r_{+}^{6}}{15}\left(1-\sqrt{1+\eta_{+}}-\frac{\Lambda}{2b^{2}}+\frac{20Q^{2}}{b^{2}\pi^{4}r_{+}^{10}}\mathcal{H}(\eta_{+}) \right)  \Bigg].
\label{eq2.13}
\end{equation}
The electrostatic potential difference between the black hole horizon and the infinity may be defined as\cite{Deh3},
\begin{equation}
\Phi=\sqrt{\frac{(n-1)}{2(n-2)}}\frac{q}{r_{+}^{n-2}}\mathcal{H}(\eta_{+})=\frac{Q}{\pi^{2}r_{+}^{4}}\mathcal{H}(\eta_{+})\qquad (\text{for} \;n= 6)
\label{eq2.15}
\end{equation}
 where 
 \begin{equation}
\eta_{+}=\dfrac{16Q^{2}}{b^{2}\pi^{4}r_{+}^{10}}.
\label{eq2.16}
\end{equation}
It is to be noted that in obtaing eq.\eqref{eq2.16} we have used eq.\eqref{eq2.8}.

The Hawking temperature for the third order LBI-AdS black hole is obtained by analytic continuation of the metric. If we set $t\rightarrow i \tau$ we obtain the Euclidean section of eq.\eqref{eq2.11} which requires to be regular at the horizon ($r_{+}$). Thus we must identify $\tau\sim \tau + \beta$, where $\beta$ ($=\frac{2\pi}{\kappa}$, $\kappa$ being the surface gravity of the black hole) is the inverse of the Hawking temperature. Therefore the Hawking temperature may be written as,
\begin{eqnarray}
\beta^{-1}=T&=&\frac{1}{4\pi}\left( \frac{\partial f(r)}{\partial r}\right) _{r_{+}}\nonumber\\
 &=&\frac{10kr_{+}^{4}+5k\alpha' r_{+}^{2}+2b^{2}r_{+}^{6}\left(1-\sqrt{1+\frac{16Q^{2}}{b^{2}\pi^{4}r_{+}^{10}}}\right)-\Lambda r_{+}^{6}}{10\pi r_{+}(r_{+}^{2}+k\alpha')^{2}}.
 \label{eq2.17}
\end{eqnarray}

It is interesting to note that, in the limit $ \alpha' \rightarrow 0 $ the corresponding expression for the Hawking temperature of the Born-Infeld AdS (BI-AdS) black hole  can be recovered as\cite{DR2},
\begin{equation}
T_{BI-AdS}=\frac{1}{4\pi}\left[ \frac{4k}{r_{+}}+\frac{4b^{2}r_{+}}{5}\left(1-\sqrt{1+\frac{Q^{2}}{b^{2}r_{+}^{4}}}\right)-\frac{2\Lambda r_{+}}{5}\right] .
\label{eq2.18}  
\end{equation} 

The entropy of the black hole may be calculated from the first law of black hole mechanics\cite{Bekenstein}. In fact, it has been found that the thermodynamic quantities (e.g. entropy, temperature, `quasilocal energy' etc.) of the LBI-AdS black holes satisfy the first law of black hole mechanics\cite{Deh1, Deh2}, \cite{Deh3}-\cite{Cai7}:
\begin{equation}
dM=TdS+\Phi dQ.
\label{eq2.19}
\end{equation}

Using eq.\eqref{eq2.19} the entropy of the black hole may be obtained as,
\begin{eqnarray}
S&=&\int_{0}^{r_{+}}\frac{1}{T}\left(\frac{\partial M}{\partial r_{+}}\right)_{Q}dr_{+}\nonumber\\
  &=&\frac{\pi^{3}}{4}\left(r_{+}^{5}+\frac{10k\alpha' r_{+}^{3}}{3}+5k^{2}\alpha'^{2}r_{+} \right)
  \label{eq2.20} .
\end{eqnarray}
where we have used eqs.\eqref{eq2.13} and \eqref{eq2.17}. At this point it is interesting to note that, identical expression for the entropy was obtained earlier using somewhat different approach\cite{Wald1}-\cite{Jacobson2}. In this approach, in an arbitrary spatial dimension $n$, the expression for the Wald entropy for higher curvature black holes is written as,
\begin{eqnarray}
S&=&\frac{1}{4}\sum_{j=1}^{[\frac{n-2}{2}]}j\alpha_{j}\int_{\mathcal{B}} d^{n-1}\omega\sqrt{|\sigma |}\,\mathcal{L}_{j-1}(\sigma)\nonumber\\
&=&\frac{1}{4}\int_{\mathcal{B}} d^{n-1}\omega\sqrt{|\sigma |}\Big(1+2\alpha_{2}\mathcal{\tilde R}+3\alpha_{3}(\tilde{R}_{abcd}\tilde{R}^{abcd}-4\tilde{R}_{ab}\tilde{R}^{ab}+
\mathcal{\bar{R}}^{2})\Big)\nonumber\\
&=&\dfrac{\mathcal{V}_{n-1}}{4}\Bigg[r_{+}^{4}+\dfrac{2(n-1)}{(n-3)}k\alpha' r_{+}^{2}+\dfrac{(n-1)}{(n-5)}k^{2}\alpha'^{2}\Bigg]r_{+}^{n-5},
\label{eq2.21}
\end{eqnarray}
$\mathcal{L}_{j}(\sigma)$ being the $j$th order Lovelock Lagrangian of $\sigma_{ij}$ and the {\it{tilde}} denotes the corresponding quantities for the induced metric $\sigma_{ij}$. If we put $n=6$ in eq.\eqref{eq2.21} we obtain the expression of eq.\eqref{eq2.20}\footnote{The entropy of black holes both in the usual Einstein gravity and in higher curvature gravity can also be obtained by using the approach of Refs.\cite{Majhi_1,Majhi_2} and \cite{B_Wang} respectively. The expression for the entropy of the third order Lovelock black hole given by eq.\eqref{eq2.20} is the same as that of Ref.\cite{B_Wang}.}. Thus, we may infer that the entropy of the black hole obtained from the first law of black hole mechanics  is indeed the Wald entropy.

From eqs.\eqref{eq2.20} and \eqref{eq2.21} we find that the entropy is not proportional to the one-fourth of the horizon area as in the case of the black holes in the Einstein gravity. However, if we take the limit $ \alpha' \rightarrow 0 $, we can recover the usual area law of black hole entropy in the BI-AdS black hole\cite{DR2} as,
\begin{equation}
S=\frac{\pi^{3}}{4}r_{+}^{5}.
\label{eq2.22}
\end{equation}

In our study of critical phenomena we will be mainly concerned about the spherically symmetric space-time. In this regard we will always take the value of $ k $ to be $ +1 $. Substituting $ k=1 $ in eqs.\eqref{eq2.17} and \eqref{eq2.20} we finally obtain the expressions for the `quasilocal energy', the Hawking temperature and the entropy of the third order LBI-AdS black hole as,
\begin{equation}
M=\frac{5\pi^{2}}{16}\Bigg[\frac{\alpha'^{2}}{3}+r_{+}^{4}+\alpha' r_{+}^{2}+\frac{2b^{2}r_{+}^{6}}{15}\left(1-\sqrt{1+\eta_{+}}-\frac{\Lambda}{2b^{2}}+\frac{20Q^{2}}{b^{2}\pi^{4}r_{+}^{10}}\mathcal{H}(\eta_{+}) \right)  \Bigg]
\label{M.M}
\end{equation}
\begin{equation}
T=\frac{10r_{+}^{4}+5\alpha' r_{+}^{2}+2b^{2}r_{+}^{6}\left(1-\sqrt{1+\frac{16Q^{2}}{b^{2}\pi^{4}r_{+}^{10}}}\right)-\Lambda r_{+}^{6}}{10\pi r_{+}(r_{+}^{2}+\alpha')^{2}}
\label{eq2.23}
\end{equation}
\begin{equation}
S=\frac{\pi^{3}}{4}\left(r_{+}^{5}+\frac{10\alpha' r_{+}^{3}}{3}+5\alpha'^{2}r_{+} \right).
\label{eq2.24}
\end{equation}
\section{Phase transition and stability of the third order LBI-AdS black hole}
In this section we aim to discuss the nature of phase transition and the stability of the third order LBI-AdS black hole. A powerful method, based on the Ehrenfest's scheme of ordinary thermodynamics, was introduced by the authors of Ref.\cite{Modak1} in order to determine the nature of phase transition in black holes. Using this analytic method, phase transition phenomena in various AdS black holes were explored\cite{Modak2}-\cite{Lala}. Also, phase transition in higher dimensional AdS black holes has been discussed in Ref.\cite{Ban3}.

In this paper we have qualitatively discussed the phase transition phenomena in the third order LBI-AdS black hole following the arguments presented in the above mentioned works. At this point it must be stressed that the method presented in Ref.\cite{Modak1} has not yet been implemented for the present black hole. However, we will not present any quantitative discussion in this regard.

\begin{figure}[h]
\begin{minipage}[b]{0.5\linewidth}
\centering
\includegraphics[angle=0,width=8cm,keepaspectratio]{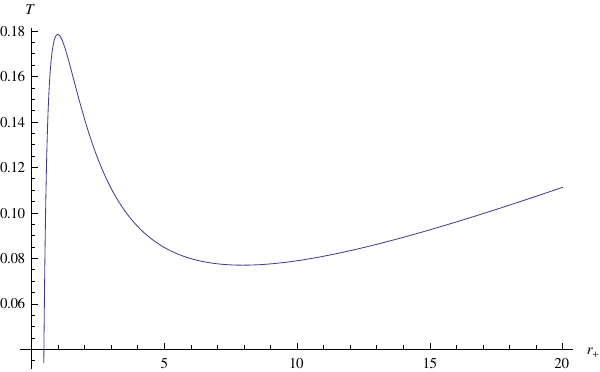}
\caption[]{\it Plot of Hawking temperature ($T $) against horizon radius ($r_{+}$),  for $ \alpha'=0.5 $, $Q=0.50$ and $ b=10 $.}
\label{Fig1}
\end{minipage}
\hspace{.1cm}
\begin{minipage}[b]{0.5\linewidth}
\centering
\includegraphics[angle=0,width=8cm,keepaspectratio]{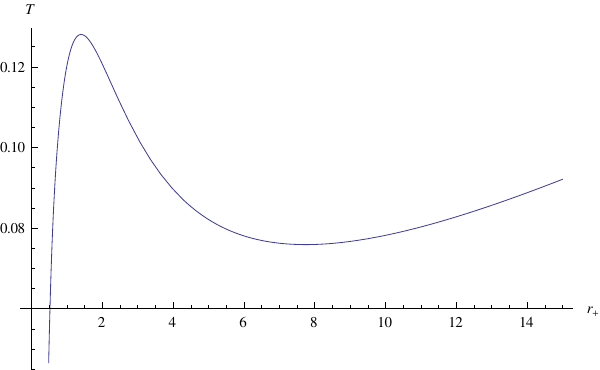}
\caption[]{\it Plot of Hawking temperature ($T $) against horizon radius ($r_{+}$),  for $ \alpha'=1.0 $, $Q=0.50$ and $ b=10 $.}
\label{Fig2}
\end{minipage}
\end{figure}

From the $T-r_{+}$ plot (Figs.\ref{Fig1},\ref{Fig2}) it is evident that there is no discontinuity in the temperature of the black hole. This rules out the possibility of first order phase transition\cite{Ban3},\cite{Modak2}-\cite{Lala}. 

In order to see whether there is any higher order phase transition, we calculate the specific heat of the black hole. In the canonical ensemble framework the specific heat at constant charge\footnote{This is analogous to the specific heat heat at constant volume ($C_{V}$) in the ordinary thermodynamics.} ($C_{Q}$) can be calculated as\cite{DR2,DR1},

\begin{equation}
C_{Q}=T\left(\frac{\partial S}{\partial T} \right)_{Q}
=T\frac{(\partial S/\partial r_{+})_{Q}}{(\partial T/\partial r_{+})_{Q}}
=\frac{\mathcal{N}(r_{+},Q)}{\mathcal{D}(r_{+},Q)}
\label{eq3.1}
\end{equation}

where 
\begin{eqnarray}
 \mathcal{N}(r_{+},Q) =\frac{5}{4}\pi^{7}r_{+}^{5}\Big(r_{+}^{2}+\alpha'\Big)^{3}\sqrt{1+\frac{16Q^{2}}{b^{2}\pi^{4}r_{+}^{10}}}\Bigg[10r_{+}^{2}+5\alpha' +2b^{2}r_{+}^{4}\Bigg(1-\sqrt{1+\frac{16Q^{2}}{b^{2}\pi^{4}r_{+}^{10}}}\Bigg)-\Lambda r_{+}^{4} \Bigg],
 \label{eq3.2}
\end{eqnarray}
\begin{eqnarray}
\mathcal{D}(r_{+},Q)&=&128Q^{2}+\Bigg[15\pi^{4}r_{+}^{6}\alpha' + 5\pi^{4}r_{+}^{4}\alpha'^{2}-\Lambda\pi^{4}r_{+}^{10}-5\pi^{4}r_{+}^{8}(2+\alpha'\Lambda)\Bigg]
\sqrt{1+\frac{16Q^{2}}{b^{2}\pi^{4}r_{+}^{10}}}\nonumber\\
&&-\Bigg(2b^{2}\pi^{4}r_{+}^{10}+10b^{2}\pi^{4}r_{+}^{8}\alpha' \Bigg) \left(1-\sqrt{1+\frac{16Q^{2}}{b^{2}\pi^{4}r_{+}^{10}}}\right).
\label{eq3.3}
\end{eqnarray}
In the derivation of eq.\eqref{eq3.1} we have used eqs.\eqref{eq2.23} and \eqref{eq2.24}.

\begin{figure}[h]
\begin{minipage}[b]{0.5\linewidth}
\centering
\includegraphics[angle=0,width=8cm,keepaspectratio]{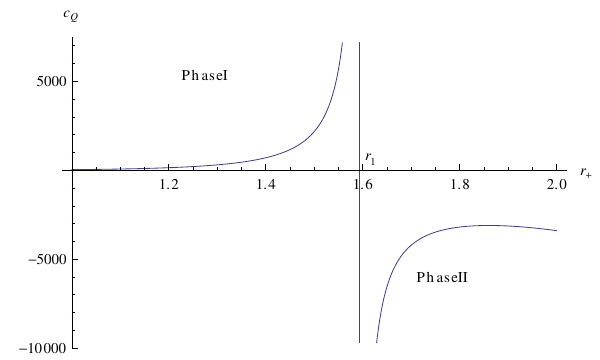}
\caption[]{\it Plot of specific heat ($C_{Q}$) against horizon radius ($r_{+}$),  for $ \alpha'=0.5 $, $Q=15$ and $ b=0.60 $ at the critical point $ r_{1} $.}
\label{Fig3}
\end{minipage}
\hspace{.1cm}
\begin{minipage}[b]{0.5\linewidth}
\centering
\includegraphics[angle=0,width=8cm,keepaspectratio]{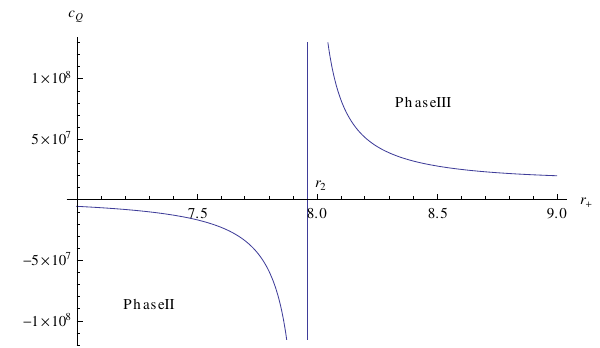}
\caption[]{\it Plot of specific heat ($C_{Q}$) against horizon radius ($r_{+}$),  for $ \alpha'=0.5 $, $Q=15$ and $ b=0.60 $ at the critical point $ r_{2} $.}
\label{Fig4}
\end{minipage}
\end{figure}

\begin{figure}[h]
\begin{minipage}[b]{0.5\linewidth}
\centering
\includegraphics[angle=0,width=8cm,keepaspectratio]{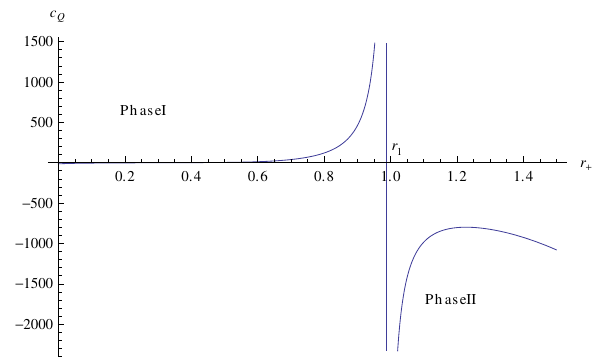}
\caption[]{\it Plot of specific heat ($C_{Q}$) against horizon radius ($r_{+}$),  for $ \alpha'=0.5 $, $Q=0.50$ and $ b=10 $ at the critical point $ r_{1} $.}
\label{Fig5}
\end{minipage}
\hspace{.1cm}
\begin{minipage}[b]{0.5\linewidth}
\centering
\includegraphics[angle=0,width=8cm,keepaspectratio]{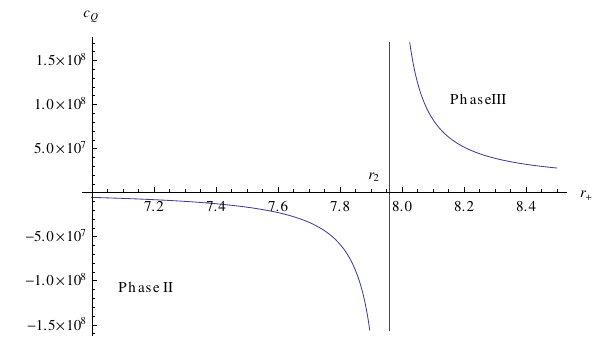}
\caption[]{\it Plot of specific heat ($C_{Q}$) against horizon radius ($r_{+}$),  for $ \alpha'=0.5 $, $Q=0.50$ and $ b=10 $ at the critical point $ r_{2} $.}
\label{Fig6}
\end{minipage}
\end{figure}

\begin{figure}[h]
\begin{minipage}[b]{0.5\linewidth}
\centering
\includegraphics[angle=0,width=8cm,keepaspectratio]{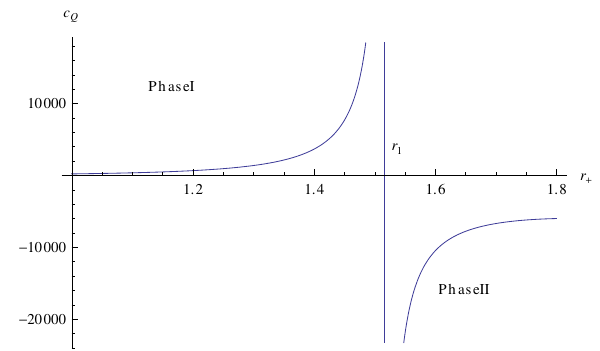}
\caption[]{\it Plot of specific heat ($C_{Q}$) against horizon radius ($r_{+}$),  for $ \alpha'=1.0 $, $Q=5$ and $ b=0.5 $ at the critical point $ r_{1} $.}
\label{Fig7}
\end{minipage}
\hspace{.1cm}
\begin{minipage}[b]{0.5\linewidth}
\centering
\includegraphics[angle=0,width=8cm,keepaspectratio]{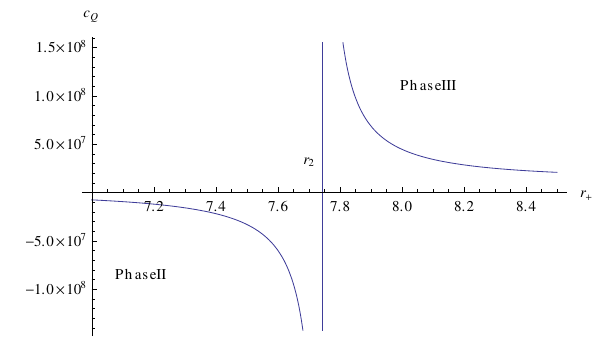}
\caption[]{\it Plot of specific heat ($C_{Q}$) against horizon radius ($r_{+}$),  for $ \alpha'=1.0 $, $Q=5$ and $ b=0.5 $ at the critical point $ r_{2} $.}
\label{Fig8}
\end{minipage}
\end{figure}

\begin{figure}[h]
\begin{minipage}[b]{0.5\linewidth}
\centering
\includegraphics[angle=0,width=8cm,keepaspectratio]{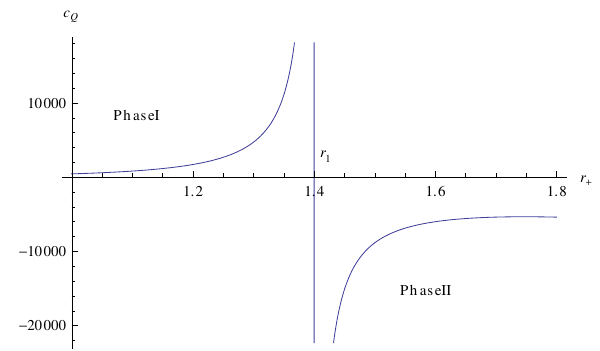}
\caption[]{\it Plot of specific heat ($C_{Q}$) against horizon radius ($r_{+}$),  for $ \alpha'=1.0 $, $Q=0.05$ and $ b=0.05 $ at the critical point $ r_{1} $.}
\label{Fig9}
\end{minipage}
\hspace{.1cm}
\begin{minipage}[b]{0.5\linewidth}
\centering
\includegraphics[angle=0,width=8cm,keepaspectratio]{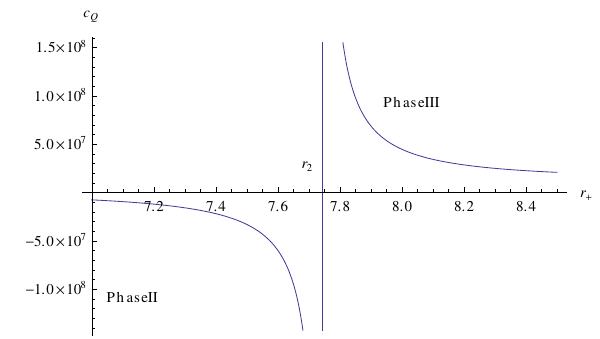}
\caption[]{\it Plot of specific heat ($C_{Q}$) against horizon radius ($r_{+}$),  for $ \alpha'=1.0 $, $Q=0.05$ and $ b=0.05 $ at the critical point $ r_{2} $.}
\label{Fig10}
\end{minipage}
\end{figure}

In Figs.\ref{Fig3} to \ref{Fig10} we have plotted $C_{Q}$ against the horizon radius $r_{+}$\footnote{Here we have zoomed in the plots near the two critical points $r_{i}$ ($i=1,2$) separately.}. The numerical values of the roots of eq.\eqref{eq3.1} are given in the Tables \ref{Table1} and \ref{Table2}. For convenience we have written the real roots of eq.\eqref{eq3.1} only. From our analysis it is observed that the specific heat always possesses simple poles. Moreover, there are two real positive roots ($r_{i}\;(i=1,2)$) of the denominator of $ C_{Q} $ for different values of the parameters $ b $, $ Q $ and $ \alpha' $. Also, from the $C_{Q}-r_{+}$ plots it is observed that the specific heat suffers discontinuity at the critical points $r_{i}\;(i=1,2)$. This property of $C_{Q}$ allows us to conclude that at the critical points there is indeed a continuous higher order phase transition\cite{DR2,DR1}.
\begin{subtables}
\begin{table}
\caption{Real roots of eq.\eqref{eq3.1} for $ \alpha'=0.5 $ and $ l=10 $}   
\centering                          
\begin{tabular}{c c c c c c c}            
\hline\hline                        
$Q$ &$ b $& $r_{1}$  & $r_{2}$ &$r_{3}$ & $r_{4} $  \\ [0.05ex]
\hline
15 & 0.6 & 1.59399 & 7.96087 & -7.96087 & -1.59399 \\
8 & 0.2 & 1.18048 & 7.96088 & -7.96088 & -1.18048 \\
5 & 0.5 & 1.25190 & 7.96088 & -7.96088 & -1.25190 \\
0.8 & 20 & 1.01695 & 7.96088 & -7.96088 & -1.01695 \\
0.3 & 15 & 0.975037 & 7.96088 & -7.96088 & -0.975037 \\
0.5 & 10 &0.989576 & 7.96088 & -7.96088 & -0.989576 \\                              
0.5 & 1 & 0.989049 & 7.96088 & -7.96088 & -0.989049 \\
0.5 & 0.5 & 0.987609 & 7.96088 & -7.96088 & -0.987609 \\
0.05 & 0.05 & 0.965701 & 7.96088 & -7.96088 & -0.965701 \\  [0.05ex]         
\hline                              
\end{tabular}\label{Table1}  
\end{table}
\begin{table}
\caption{Real roots of eq.\eqref{eq3.1} for $ \alpha'=1.0 $ and $ l=10 $}   
\centering                          
\begin{tabular}{c c c c c c c}            
\hline\hline                        
$Q$ &$ b $& $r_{1}$  & $r_{2}$ &$r_{3}$ & $r_{4} $  \\ [0.05ex]
\hline
15 & 0.6 & 1.76824 & 7.74534 & -7.74534 & -1.76824 \\
8 & 0.2 & 1.53178 & 7.74535 & -7.74535 & -1.53178 \\
5 & 0.5 & 1.51668 & 7.74535 & -7.74535 & -1.51668 \\
0.8 & 20 & 1.40553 & 7.74535 & -7.74535 & -1.40553 \\
0.3 & 15 & 1.40071 & 7.74535 & -7.74535 & -1.40071 \\
0.5 & 10 & 1.40214 & 7.74535 & -7.74535 & -1.40214 \\                              
0.5 & 1 & 1.40214 & 7.74535 & -7.74535 & -1.40214 \\
0.5 & 0.5 & 1.40213 & 7.74535 & -7.74535 & -1.40213 \\
0.05 & 0.05 & 1.39992 & 7.74535 & -7.74535 & -1.39992 \\  [0.05ex]         
\hline                              
\end{tabular}\label{Table2}  
\end{table}
\end{subtables}

Let us now see whether there is any bound in the values of the parameters $b,Q$ and $\alpha'$. At this point it must be stressed that a bound in the parameter values ($b,Q$) for the Born-Infeld-AdS black holes in ($3+1$)-dimensions was found earlier\cite{DR1,Myung3}. Moreover, this bound is removed if we consider space-time dimensions greater than four\cite{DR2}. Thus, it will be very much interesting to check whether the third order LBI-AdS black holes possess similar features. In order to do so, we will consider the extremal third order LBI-AdS black hole. In this case both $ f(r) $ and $ \frac{df}{dr} $ vanish at the degenerate horizon $ r_{e} $\cite{DR2,DR1,Myung3}. The above two conditions for extremality results the following equation:
\begin{equation}
10r_{e}^{4}+5\alpha' r_{e}^{2}+2b^{2}r_{e}^{6}\left(1-\sqrt{1+\frac{16Q^{2}}{b^{2}\pi^{4}r_{e}^{10}}}\right)-\Lambda r_{e}^{6}=0.
\label{eq3.4}
\end{equation}

In Tables \ref{Table3} and \ref{Table4} we give the numerical solutions of eq.\eqref{eq3.4} for different choices of the values of the parameters $ b $ and $ Q $ for fixed values of $ \alpha' $. From these analysis we observe that for arbitrary choices of the parameters $ b $ and $ Q $ we always obtain atleast one real positive root of eq.\eqref{eq3.4}. This implies that there exists a smooth extremal limit for arbitrary $ b $ and $ Q $ and there is no bound on the parameter space for a particular value of $ \alpha' $. Thus, the result obtained here (regarding the bound in the parameter values) is in good agreement with that obtained in Ref.\cite{DR2}.
\begin{subtables}
\begin{table}
\caption{Roots of eq.\eqref{eq3.4} for $ \alpha'=0.5 $ and $ l=10 $}   
\centering                          
\begin{tabular}{c c c c c c c}            
\hline\hline                        
$Q$ &$ b $& $r_{e1}^{2}$  & $r_{e2}^{2}$ &$r_{e3,e4}^{2}$ & $r_{e5}^{2} $  \\ [0.05ex]
\hline
15 & 0.6 & -66.4157 & - & - & +0.632734 \\
8 & 0.2 & -66.4157 & - & - & +0.121590 \\
5 & 0.5 & -66.4157 & - & - & +0.200244 \\
0.8 & 20 & -66.4157 & -0.408088 & -0.05881$ \pm i $0.304681 & +0.268514 \\                        
0.3 & 15 & -66.4157 & -0.304225 & -0.0517567$ \pm i $0.175488 & +0.145685 \\
0.5 & 10 & -66.4157 & -0.349764 & -0.0593449$ \pm i $0.240144 & +0.192519 \\
0.5 & 1 & -66.4157 & - & - & +0.0221585 \\                              
0.5 & 0.5 & -66.4157 & - & - & +0.00625335 \\ 
0.05 & 0.05 &-66.4157 & - & - & +6.57019$ \times $10$^{-7}$ \\  [0.05ex]         
\hline                              
\end{tabular}\label{Table3}  
\end{table}
\begin{table}
\caption{Roots of eq.\eqref{eq3.4} for $ \alpha'=1.0 $ and $ l=10 $}   
\centering                          
\begin{tabular}{c c c c c c c}            
\hline\hline                        
$Q$ &$ b $& $r_{e1}^{2}$  & $r_{e2}^{2}$ &$r_{e3,e4}^{2}$ & $r_{e5}^{2} $  \\ [0.05ex]
\hline
15 & 0.6 & -66.1628 & - & - & +0.50473 \\
8 & 0.2 & -66.1628 & - & - & +0.0546589 \\
5 & 0.5 & -66.1628 & - & - & +0.110054 \\
0.8 & 20 & -66.1629 & -0.563564 & -0.0913395$ \pm i $0.266887 & +0.236186 \\                        
0.3 & 15 & -66.1629 & -0.514815 & -0.0584676$ \pm i $0.14609 & +0.116834 \\
0.5 & 10 & -66.1629 & -0.531635 & -0.0760937$ \pm i $0.210702 & +0.155024 \\
0.5 & 1 & -66.1629 & -0.542417 & - & +0.00640486 \\                              
0.5 & 0.5 & -66.1629 & - & - & +0.00163189 \\ 
0.05 & 0.05 & -66.1629 & - & - & +1.64256$ \times $10$^{-7}$ \\  [0.05ex]           
\hline                              
\end{tabular}\label{Table4}  
\end{table}
\end{subtables}


We shall now analyse the thermodynamic stability of the third order LBI-AdS black hole. This is generally done by studying the behaviour of $ C_{Q} $ at the critical points\cite{DR2, Ban1, Lala, DR1, Myung3}. The $ C_{Q}-r_{+} $ plots show that there are indeed three phases of the black hole. These phases can be classified as - Phase I ($ 0<r_{+}<r_{1} $), Phase II ($ r_{1}<r_{+}<r_{2} $) and Phase III ($ r_{+}>r_{2} $). Since the higher mass black hole possesses larger entropy/horizon radius, there is a phase transition at $ r_{1} $ from smaller mass black hole (Phase I) to intermediate (higher mass) black hole (Phase II). The critical point $ r_{2} $ corresponds to a phase transition from an intermediate (higher mass) black hole (Phase II) to a larger mass black hole (Phase III). Moreover, from the $ C_{Q}-r_{+} $ plots we note that the specific heat $ C_{Q} $ is positive for Phase I and Phase III whereas it is negative for Phase II. Therefore Phase I and Phase III correspond to thermodynamically stable phase ($ C_{Q}>0 $) whereas Phase II corresponds to thermodynamically unstable phase ($ C_{Q}<0 $).

We can further extend our stability analysis by considering the free energy of the third order LBI-AdS black hole. The free energy plays an important role in the theory of phase transition and critical phenomena. We may define the free energy of the third order LBI-AdS black hole as,
\begin{equation}
\mathcal{F}(r_{+},Q)=M(r_{+},Q)-TS.
\label{eq3.5}
\end{equation}

Using eqs.\eqref{M.M}, \eqref{eq2.23} and \eqref{eq2.24} we can write eq.\eqref{eq3.5} as,
\begin{eqnarray}
\mathcal{F}&=&\frac{5\pi^{2}}{16}\left[\frac{\alpha'^{2}}{3}+r_{+}^{4}+\alpha' r_{+}^{2}+\frac{2b^{2}r_{+}^{6}}{15}\left(1-\sqrt{1+\frac{16Q^{2}}{b^{2}\pi^{4}r_{+}^{10}}}-\frac{\Lambda}{2b^{2}}+\frac{20Q^{2}}{b^{2}\pi^{4}r_{+}^{10}}\mathcal{H}\left( \frac{1}{2},\frac{2}{5},\frac{7}{5},-\frac{16Q^{2}}{b^{2}\pi^{4}r_{+}^{10}}\right)  \right)  \right] \nonumber\\
&&-\frac{\pi^{2}(r_{+}^{5}+\frac{10\alpha' r_{+}^{3}}{3}+5\alpha'^{2}r_{+})}{40r_{+}(r_{+}^{2}+\alpha')^{2}}\left[10r_{+}^{4}+5\alpha' r_{+}^{2}+2b^{2}r_{+}^{6}\left(1-\sqrt{1+\frac{16Q^{2}}{b^{2}\pi^{4}r_{+}^{10}}}\right)-\Lambda r_{+}^{6}  \right]. 
\label{eq3.6}
\end{eqnarray}

\begin{figure}[h]
\begin{minipage}[b]{0.5\linewidth}
\centering
\includegraphics[angle=0,width=8cm,keepaspectratio]{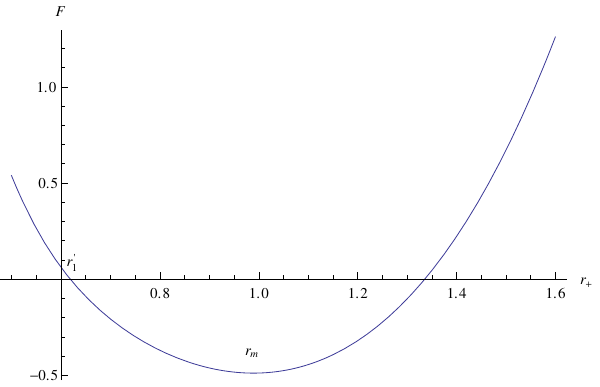}
\caption[]{\it Plot of free energy($\mathcal{F}$) against horizon radius ($r_{+}$),  for $ \alpha'=0.50 $, $Q=0.50$ and $ b=10 $.}
\label{Fig11}
\end{minipage}
\hspace{.1cm}
\begin{minipage}[b]{0.5\linewidth}
\centering
\includegraphics[angle=0,width=8cm,keepaspectratio]{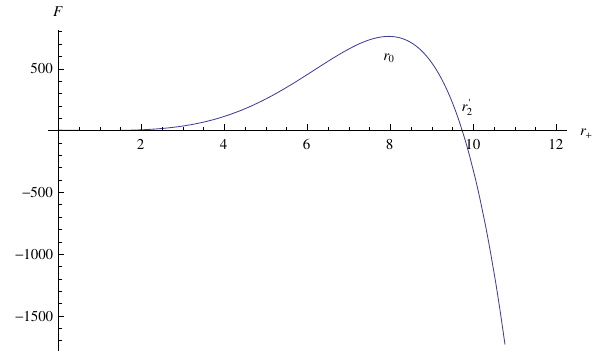}
\caption[]{\it Plot of free energy($\mathcal{F}$) against horizon radius ($r_{+}$),  for $ \alpha'=0.50 $, $Q=0.50$ and $ b=10 $.}
\label{Fig12}
\end{minipage}
\end{figure}

\begin{figure}[h]
\begin{minipage}[b]{0.5\linewidth}
\centering
\includegraphics[angle=0,width=8cm,keepaspectratio]{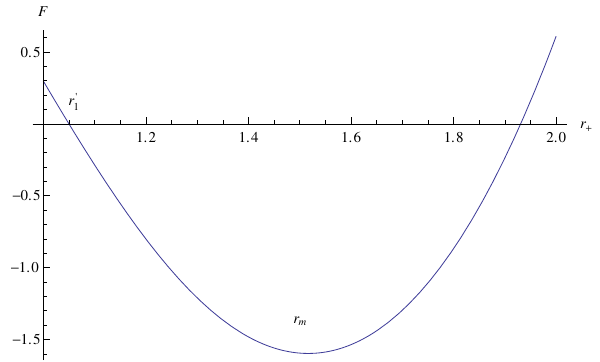}
\caption[]{\it Plot of free energy($\mathcal{F}$) against horizon radius ($r_{+}$),  for $ \alpha'=1.0 $, $Q=5$ and $ b=0.5 $.}
\label{Fig13}
\end{minipage}
\hspace{.1cm}
\begin{minipage}[b]{0.5\linewidth}
\centering
\includegraphics[angle=0,width=8cm,keepaspectratio]{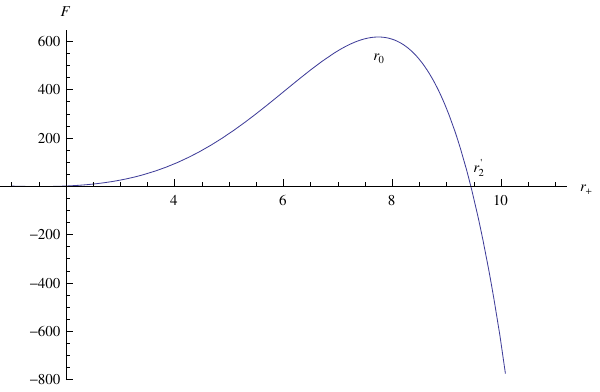}
\caption[]{\it Plot of free energy($\mathcal{F}$) against horizon radius ($r_{+}$),  for $ \alpha'=1.0 $, $Q=5$ and $ b=0.5 $.}
\label{Fig14}
\end{minipage}
\end{figure}

In Figs.\ref{Fig11} to \ref{Fig14} we have given the plots of the free energy ($ \mathcal{F} $) of the black hole with the radius of the outer horizon $ r_{+} $.  The free energy ($ \mathcal{F} $) has a minima $ \mathcal{F}=\mathcal{F}_{m} $ at $ r_{+}=r_{m} $. This point of minimum free energy is exactly the same as the first critical point $ r_{+}=r_{1} $ where the black hole shifts from a stable to an unstable phase. On the other hand $ \mathcal{F} $ has a maxima $ \mathcal{F}=\mathcal{F}_{0} $ at $ r_{+}=r_{0} $. The point at which $ \mathcal{F} $ reaches its maximum value, is identical with the second critical point $ r_{+}=r_{2} $ where the black hole changes from unstable to stable phase. We can further divide the $ \mathcal{F}-r_{+} $ plot into three distinct regions. In the first region $ r_{1}^{\prime}<r_{+}<r_{m} $ the negative free energy decreases until it reaches the minimum value ($ \mathcal{F}_{m} $) at $ r_{+}=r_{m} $. This region corresponds to the stable phase (Phase I: $ C_{Q}>0 $) of the black hole. The free energy changes its slope at $ r_{+}=r_{m} $ and continues to increase in the  second region $ r_{m}<r_{+}<r_{0} $ approaching towards the maximum value ($ \mathcal{F}_{0} $) at $ r_{+}=r_{0} $. This region corresponds to the Phase II of the $ C_{Q}-r_{+} $ plot where the black hole becomes unstable ($ C_{Q}<0 $). The free energy changes its slope once again at $ r_{+}=r_{0} $ and decreases to zero at $ r_{+}=r_{2}^{\prime} $ and finally becomes negative for $ r_{+}>r_{2}^{\prime} $. This region of the $ \mathcal{F}-r_{+} $ plot corresponds to the Phase III of the $ C_{Q}-r_{+} $ plot where the black hole finally becomes stable ($ C_{Q}>0 $).


\section{Critical exponents and scaling hypothesis }
In thermodynamics, the theory of phase transition plays a crucial role to understand the behavior of a thermodynamic system. The behavior of thermodynamic quantities near the critical point(s) of phase transition gives a considerable amount of information about the system. The behavior of a thermodynamic system near the critical point(s) is usually studied by means of a set of indices known as the critical exponents\cite{Hankey,Stanley}. These are generally denoted by a set of Greek letters: $ \alpha,\,\beta,\,\gamma,\,\delta,\,\phi,\,\psi,\,\eta,\,\nu $. The critical exponents describe the nature of singularities in various measurable thermodynamic quantities near the critical point(s). 

In this section we aim to determine the first six static critical exponents ($ \alpha,\,\beta,\,\gamma,\,\delta,\,\phi,\,\psi $). For this purpose we shall follow the method discussed in Refs.\cite{DR2, Bibhas, DR1}. We shall then discuss about the static scaling laws and static scaling hypothesis. We shall determine the other two critical exponents ($ \nu $ and $ \eta $) from two additional scaling laws.

\begin{flushleft}
\textit{\textbf{Critical exponent $ \alpha $:}}
\end{flushleft}
In order to determine the critical exponent $ \alpha $ which is associated with the singularity of $ C_{Q} $ near the critical points $ r_{i} $ ($ i=1,\,2 $), we choose a point in the infinitesimal neighborhood of $ r_{i} $ as,
\begin{equation}
r_{+}=r_{i}(1+\Delta),\,\,\,\,\,\, i=1,\,2
\label{eq4.1}
\end{equation}
where $ |\Delta|<<1 $. Let us denote the temperature at the critical point by $ T(r_{i}) $ and define the quantity
\begin{equation}
\epsilon = \frac{T(r_{+})-T(r_{i})}{T(r_{i})}
\label{eq4.2}
\end{equation}
such that $ |\epsilon|<<1 $.

We now Taylor expand $ T(r_{+}) $ in the neighborhood of $ r_{i} $ keeping the charge constant ($ Q=Q_{c} $), which yields,
\begin{equation}
T(r_{+})=T(r_{i})+\left[\left(\frac{\partial T}{\partial r_{+}} \right)_{Q=Q_{c}}  \right]_{r_{+}=r_{i}}(r_{+}-r_{i})+\frac{1}{2} \left[\left(\frac{\partial^{2} T}{\partial r_{+}^{2}} \right)_{Q=Q_{c}}  \right]_{r_{+}=r_{i}}(r_{+}-r_{i})^{2}+........
\label{eq4.3}
\end{equation}
Since the divergence of $ C_{Q} $ results from the vanishing of $ \left(\frac{\partial T}{\partial r_{+}} \right)_{Q}$ at the critical point $ r_{i} $ (eq.\eqref{eq3.1}), we may write eq.\eqref{eq4.3} as,
\begin{equation}
T(r_{+})=T(r_{i})+\frac{1}{2} \left[\left(\frac{\partial^{2} T}{\partial r_{+}^{2}} \right)_{Q=Q_{c}}  \right]_{r_{+}=r_{i}}(r_{+}-r_{i})^{2}
\label{eq4.4}
\end{equation}
where we have neglected the higher order terms in eq.\eqref{eq4.3}.

Using eqs.\eqref{eq4.1} and \eqref{eq4.2} we can finally write eq.\eqref{eq4.4} as,
\begin{equation}
\Delta = \frac{\epsilon^{1/2}}{\Gamma_{i}^{1/2}}
\label{eq4.5}
\end{equation}

where \begin{equation}
\Gamma_{i}=\frac{r_{i}^{2}}{2T(r_{i})} \left[\left(\frac{\partial^{2} T}{\partial r_{+}^{2}} \right)_{Q=Q_{c}}  \right]_{r_{+}=r_{i}}.
\label{eq4.6}
\end{equation}

The detailed expression of $ \Gamma_{i} $ is very much cumbersome and we shall not write it for the present work.

If we examine the $ T-r_{+} $ plots (Figs. \ref{Fig1} and \ref{Fig2}) we observe that, near the critical point $r_{+}=r_{1} $ (which corresponds to the `hump') $ T(r_{+})<T(r_{1}) $ so that $ \epsilon <0 $ and on the contrary, near the critical point $r_{+}=r_{2} $ (which corresponds to the `dip') $ T(r_{+})>T(r_{2}) $ implying $ \epsilon >0 $.

Substituting eq.\eqref{eq4.1} into eq.\eqref{eq3.1} we can write the singular part of $ C_{Q} $ as,
\begin{equation}
C_{Q}=\frac{\mathcal{N}^{\prime}(r_{i},Q_{c})}{\Delta\cdot \mathcal{D}^{\prime}(r_{i},Q_{c})}
\label{eq4.7}
\end{equation}

where $ \mathcal{N}^{\prime}(r_{i},Q_{c}) $ is the value of the numerator of $ C_{Q} $  (eq.\eqref{eq3.2}) at the critical point $ r_{+}=r_{i} $ and critical charge $ Q=Q_{c} $. The expression for $ \mathcal{D}^{\prime}(r_{i},Q_{c}) $ is given by,
\begin{eqnarray}
\mathcal{D}^{\prime}(r_{i},Q_{c})=\mathcal{D}_{1}^{\prime}(r_{i},Q_{c})+\mathcal{D}_{2}^{\prime}(r_{i},Q_{c})+\mathcal{D}_{3}^{\prime}(r_{i},Q_{c})
\label{eq4.8}
\end{eqnarray}
where 
\begin{eqnarray*}
\mathcal{D}_{1}^{\prime}(r_{i},Q_{c})&=&10\pi^{4}r_{i}^{4}\sqrt{1+\frac{16Q_{c}^{2}}{b^{2}\pi^{4}r_{i}^{10}}}\Bigg[\Big(2\alpha'^{2}+2b^{2}r_{i}^{6}+8b^{2}r_{i}^{4}\alpha'\Big)-\Big(\Lambda r_{i}^{6}+9\alpha' r_{i}^{2}+4(2+\Lambda\alpha')r_{i}^{4}\Big)\Bigg]\\
\mathcal{D}_{2}^{\prime}(r_{i},Q_{c})&=&-\frac{80Q_{c}^{2}}{b^{2}r_{i}^{2}}\left[\frac{15\alpha'}{r_{i}^{2}}+\frac{5\alpha'^{2}}{r_{i}^{4}}-\Lambda r_{i}^{2}-5(2+\Lambda\alpha') \right] \\
\mathcal{D}_{3}^{\prime}(r_{i},Q_{c})&=&-20b^{2}\pi^{4}r_{i}^{8}\left[r_{i}^{2}\left(1+\frac{8Q_{c}^{2}}{b^{2}\pi^{4}r_{i}^{10}}\right)+4\alpha'\left(1+\frac{10Q_{c}^{2}}{b^{2}\pi^{4}r_{i}^{10}}\right)\right]. 
\end{eqnarray*}


It is to be noted that while expanding the denominator of $ C_{Q} $ we have retained the terms which are linear in $ \Delta $ and all other higher order terms of $ \Delta $ have been neglected.

Using eq.\eqref{eq4.7} we may summarize the critical behavior of $ C_{Q} $ near the critical points ($ r_{1} $ and $ r_{2} $) as follows:
\begin{center}
 \begin{equation}
 C_{Q}\sim
\left\{
\begin{array}{lr}
\left[\frac{\mathcal{A}_{i}}{(-\epsilon)^{1/2}} \right]_{r_{i}=r_{1}}&\epsilon<0 \\
\left[\frac{\mathcal{A}_{i}}{(+\epsilon)^{1/2}} \right]_{r_{i}=r_{2}}&\epsilon>0                              
\end{array}
\right.
\label{eq4.9}
\end{equation}
\end{center}

where 
\begin{equation}
\mathcal{A}_{i}=\dfrac{\Gamma_{i}^{1/2}\mathcal{N}^{\prime}(r_{i},Q_{c})}{\mathcal{D}^{\prime}(r_{i},Q_{c})}.
\label{eq4.10}
\end{equation}

We can combine the r.h.s of eq.\eqref{eq4.9} into a single expression, which describes the singular nature of $ C_{Q} $ near the critical point $ r_{i} $, yielding
\begin{eqnarray}
C_{Q}&=&\frac{\mathcal{A}_{i}}{|\epsilon|^{1/2}}\nonumber\\
         &=&\frac{\mathcal{A}_{i}T_{i}^{1/2}}{|T-T_{i}|^{1/2}}.
         \label{eq4.11}
\end{eqnarray}
where we have used eq.\eqref{eq4.2}. Here $ T $ and $ T_{i} $ are the abbreviations of $ T(r_{+}) $ and $ T(r_{i}) $ respectively.

We can now compare eq.\eqref{eq4.11} with the standard form 
\begin{equation}
C_{Q}\sim |T-T_{i}|^{-\alpha}
\label{eq4.12}
\end{equation}
which gives $ \alpha =\dfrac{1}{2} $.

\begin{flushleft}
\textit{\textbf{Critical exponent $\beta$ :}}
\end{flushleft}

The critical exponent $ \beta  $ is related to the electric potential at infinity ($ \Phi $) by the relation
\begin{equation}
\Phi (r_{+})-\Phi (r_{i})\sim |T-T_{i}|^{\beta}
\label{eq4.13}
\end{equation}
where the charge ($Q$) is kept constant.

Near the critical point $ r_{+}=r_{i} $ the Taylor expansion of $ \Phi (r_{+}) $ yields,
\begin{equation}
\Phi (r_{+})=\Phi (r_{i})+\left[\left(\frac{\partial \Phi}{\partial r_{+}} \right)_{Q=Q_{c}}  \right]_{r_{+}=r_{i}}(r_{+}-r_{i})+...... 
\label{eq4.14}
\end{equation}

Neglecting the higher order terms and using eqs.\eqref{eq2.15} and \eqref{eq4.5} we may rewrite eq.\eqref{eq4.14} as,
\begin{equation}
\Phi (r_{+})-\Phi (r_{i})=-\left( \frac{4Q_{c}}{\pi^{2}r_{i}^{4}\Gamma_{i}^{1/2}T_{i}^{1/2}\sqrt{1+\frac{16Q_{c}^{2}}{b^{2}\pi^{4}r_{i}^{10}}}}\right) |T-T_{i}|^{1/2}.
\label{eq4.15}
\end{equation}

Comparing eq.\eqref{eq4.15} with eq.\eqref{eq4.13} we finally obtain $ \beta=\dfrac{1}{2} $.

\begin{flushleft}
\textit{\textbf{Critical exponent $\gamma$}:}
\end{flushleft}

We shall now determine the critical exponent $\gamma$ which is associated with the singularity of the inverse of the isothermal compressibility ($\kappa_{T}^{-1}$) at constant charge $ Q=Q_{c} $ near the critical point $ r_{+}=r_{i} $ as,
\begin{equation}
\kappa_{T}^{-1}\sim |T-T_{i}|^{-\gamma}.
\label{eq4.16}
\end{equation}

In order to calculate $ \kappa_{T}^{-1} $ we use the standard thermodynamic definition
\begin{eqnarray}
\kappa_{T}^{-1}&=&Q\left(\frac{\partial \Phi}{\partial Q} \right)_{T}\nonumber\\
                          &=&-Q\left(\frac{\partial \Phi}{\partial T} \right)_{Q}\left(\frac{\partial T}{\partial Q} \right)_{\Phi}
\label{eq4.17}                          
\end{eqnarray}

where in the last line of eq.\eqref{eq4.17} we have used the identity
\begin{equation}
\left(\frac{\partial \Phi}{\partial T} \right)_{Q}\left(\frac{\partial T}{\partial Q} \right)_{\Phi}\left(\frac{\partial Q}{\partial \Phi} \right)_{T}=-1
\label{eq4.18}
\end{equation}

Using eqs.\eqref{eq2.15} and \eqref{eq2.23} we can write eq.\eqref{eq4.17} as,
\begin{equation}
\kappa_{T}^{-1}=\dfrac{\Omega(r_{+},Q)}{\mathcal{D}(r_{+},Q)}
\label{eq4.19}
\end{equation}

where $ \mathcal{D}(r_{+},Q) $ is the denominator identically equal to eq.\eqref{eq3.3} (the denominator of $ C_{Q} $) and the expression for $ \Omega(r_{+},Q) $ may be written as,
\begin{eqnarray}
\Omega(r_{+},Q)&=&\frac{Q}{5\pi^{2}r_{+}^{4}}\Bigg[128Q^{2}+\Big(15\pi^{4}r_{+}^{6}\alpha +5\pi^{4}r_{+}^{4}\alpha^{2}-\Lambda\pi^{4}r_{+}^{10}-5\pi^{4}r_{+}^{8}(2+\alpha\Lambda)\Big)\nonumber\\
&&\sqrt{1+\frac{16Q^{2}}{b^{2}\pi^{4}r_{+}^{10}}}-\Big(2b^{2}\pi^{4}r_{+}^{10}+10b^{2}\pi^{4}r_{+}^{8}\alpha\Big)
\Bigg(1-\sqrt{1+\frac{16Q^{2}}{b^{2}\pi^{4}r_{+}^{10}}}\Bigg)\nonumber\\
&&\mathcal{H}\Bigg(\frac{1}{2},\frac{2}{5},\frac{7}{5},-\frac{16Q^{2}}{b^{2}\pi^{4}r_{+}^{10}}\Bigg)\Bigg]-\Sigma(r_{+}, Q)
\label{eq4.20}
\end{eqnarray}
where
\begin{eqnarray*}
\Sigma(r_{+}, Q)&=&\frac{4Q}{5}\Bigg[2b^{2}\pi^{2}r_{+}^{4}(r_{+}^{2}+5\alpha)\sqrt{1+\frac{16Q^{2}}{b^{2}\pi^{4}r_{+}^{10}}}+\pi^{2}\Big(-15r^{2}_{+}\alpha -5\alpha^{2}+r_{+}^{6}(\Lambda -2b^{2})\nonumber\\
&&+5r_{+}^{4}(2+\Lambda\alpha -2b^{2}\alpha)\Big)\Bigg]
\end{eqnarray*}

From eq.\eqref{eq4.19} we observe that $ \kappa_{T}^{-1} $ possesses simple poles. Moreover $ \kappa_{T}^{-1} $ and $ C_{Q} $ exhibit common singularities.

We are now interested in the behavior of $ \kappa_{T}^{-1} $ near the critical point $ r_{+}=r_{i} $. In order to do so we substitute eq.\eqref{eq4.1} into eq.\eqref{eq4.19}. The resulting equation for the singular part of $ \kappa_{T}^{-1} $ may be written as,
\begin{equation}
\kappa_{T}^{-1}=\dfrac{\Omega^{\prime}(r_{i},Q_{c})}{\Delta\cdot \mathcal{D}^{\prime}(r_{i},Q_{c})}.
\label{eq4.21}
\end{equation}

In eq.\eqref{eq4.21}, $ \Omega^{\prime}(r_{i},Q_{c}) $ is the value of the numerator of $ \kappa_{T}^{-1} $ (eq.\eqref{eq4.20}) at the critical point $r_{+}=r_{i} $ and critical charge $ Q=Q_{c} $. Whereas $ \mathcal{D}^{\prime}(r_{i},Q_{c}) $ was identified earlier (eq.\eqref{eq4.8}).

Substituting eq.\eqref{eq4.5} in eq.\eqref{eq4.21} we may express the singular nature of $ \kappa_{T}^{-1} $ near the critical points ($ r_{1} $ and $ r_{2} $) as,
\begin{center}
 \begin{equation}
 \kappa_{T}^{-1}\simeq
\left\{
\begin{array}{lr}
\left[\frac{\mathcal{B}_{i}}{(-\epsilon)^{1/2}} \right]_{r_{i}=r_{1}}&\epsilon<0 \\
\left[\frac{\mathcal{B}_{i}}{(+\epsilon)^{1/2}} \right]_{r_{i}=r_{2}}&\epsilon>0                              
\end{array}
\right.
\label{eq4.22}
\end{equation}
\end{center}

where
\begin{equation}
\mathcal{B}_{i}=\dfrac{\Gamma_{i}^{1/2}\Omega^{\prime}(r_{i},Q_{c})}{ \mathcal{D}^{\prime}(r_{i},Q_{c})}
\label{eq4.23}.
\end{equation}

Combining the r.h.s of eq.\eqref{eq4.22} into a single expression as before, we can express the singular behavior of $ \kappa_{T}^{-1} $ near the critical point $ r_{i} $ as,
\begin{eqnarray}
\kappa_{T}^{-1}&=&\frac{\mathcal{B}_{i}}{|\epsilon|^{1/2}}\nonumber\\
         &=&\frac{\mathcal{B}_{i}T_{i}^{1/2}}{|T-T_{i}|^{1/2}}.
\label{eq4.24}         
\end{eqnarray}

Comparing eq.\eqref{eq4.24} with eq.\eqref{eq4.16} we find $ \gamma=\dfrac{1}{2} $.
\begin{flushleft}
\textit{\textbf{Critical exponent $\delta$}:}
\end{flushleft}

Let us now calculate the critical exponent $ \delta $ which is associated with the electrostatic potential ($ \Phi $) for the fixed value $ T=T_{i} $ of temperature. The relation can be written as,
\begin{equation}
\Phi (r_{+})-\Phi (r_{i})\sim |Q-Q_{i}|^{1/\delta}.
\label{eq4.25}
\end{equation} 
 In this relation $ Q_{i} $ is the value of charge ($ Q $) at the critical point $ r_{i} $. In order to obtain $ \delta $ we first Taylor expand $ Q(r_{+}) $ around the critical point $ r_{+}=r_{i} $. This yields,
 \begin{equation}
Q(r_{+})=Q(r_{i})+\left[\left(\frac{\partial Q}{\partial r_{+}} \right)_{T=T_{i}}  \right]_{r_{+}=r_{i}}(r_{+}-r_{i})+\frac{1}{2} \left[\left(\frac{\partial^{2} Q}{\partial r^{2}_{+}} \right)_{T=T_{i}}  \right]_{r_{+}=r_{i}}(r_{+}-r_{i})^{2}+.....
\label{eq4.26}
\end{equation}

Neglecting the higher order terms we can write eq.\eqref{eq4.26} as,
\begin{equation}
Q(r_{+})-Q(r_{i})=\frac{1}{2} \left[\left(\frac{\partial^{2} Q}{\partial r^{2}_{+}} \right)_{T}  \right]_{r_{+}=r_{i}}(r_{+}-r_{i})^{2}
\label{eq4.27}
\end{equation}
 Here we have used the standard thermodynamic identity
 \begin{equation}
\left[\left( \frac{\partial Q}{\partial r_{+}}\right)_{T}  \right]_{r_{+}=r_{i}}\left[\left( \frac{\partial r_{+}}{\partial T}\right)_{Q} \right]_{r_{+}=r_{i}} \left(\frac{\partial T}{\partial Q}\right)_{r_{+}=r_{i}}=-1
\label{eq4.28}
\end{equation}
and considered the fact that at the critical point $ r_{+}=r_{i} $, $ \left( \frac{\partial T}{\partial r_{+}}\right)_{Q} $ vanishes.

Let us now define a quantity
\begin{equation}
\Upsilon = \frac{Q(r_{+})-Q_{i}}{Q_{i}}=\frac{Q-Q_{i}}{Q_{i}}
\label{eq4.29}
\end{equation}
where $ |\Upsilon|<<1 $. Here we denote $ Q(r_{+}) $ and $ Q(r_{i}) $ by $ Q $ and $ Q_{i} $ respectively.

Using eqs.\eqref{eq4.1} and \eqref{eq4.29} we obtain from eq.\eqref{eq4.27} 
\begin{equation}
\Delta=\frac{\Upsilon^{1/2}}{\Psi^{1/2}_{i}}\left[ \frac{2Q_{i}}{r_{i}^{2}}\right]^{1/2} 
\label{eq4.30}
\end{equation}
where
\begin{equation}
\Psi_{i}= \left[\left(\frac{\partial^{2} Q}{\partial r^{2}_{+}} \right)_{T}  \right]_{r_{+}=r_{i}}.
\label{eq4.31}
\end{equation}

The expression for $ \Psi_{i} $ is very much cumbersome and we shall not write it here.

We shall now consider the functional relation 
\begin{equation}
\Phi = \Phi(r_{+},Q)
\label{eq4.32}
\end{equation}
from which we may write,
\begin{equation}
\left[ \left(\frac{\partial \Phi}{\partial r_{+}} \right)_{T} \right]_{r_{+}=r_{i}}= \left[ \left(\frac{\partial \Phi}{\partial r_{+}} \right)_{Q} \right]_{r_{+}=r_{i}}+\left[ \left(\frac{\partial Q}{\partial r_{+}} \right)_{T} \right]_{r_{+}=r_{i}}\left(\frac{\partial \Phi}{\partial Q} \right)_{r_{+}=r_{i}}. 
\label{eq4.33}
\end{equation}

Using eq.\eqref{eq4.28} we can rewrite eq.\eqref{eq4.33} as,
\begin{equation}
\left[ \left(\frac{\partial \Phi}{\partial r_{+}} \right)_{T=T_{i}} \right]_{r_{+}=r_{i}}= \left[ \left(\frac{\partial \Phi}{\partial r_{+}} \right)_{Q=Q_{c}} \right]_{r_{+}=r_{i}}.
\label{eq4.34}
\end{equation}

Now the Taylor expansion of $ \Phi $ at constant temperature around $ r_{+}=r_{i} $ yields,
\begin{equation}
\Phi (r_{+})=\Phi (r_{i})+\left[ \left(\frac{\partial \Phi}{\partial r_{+}} \right)_{T=T_{i}} \right]_{r_{+}=r_{i}}(r_{+}-r_{i})
\label{eq4.35}
\end{equation}
where we have neglected all the higher order terms.

Finally using eqs.\eqref{eq4.30}, \eqref{eq4.34} and \eqref{eq2.15} we may write eq.\eqref{eq4.35} as,
\begin{equation}
\Phi (r_{+})-\Phi (r_{i})=\left( \frac{-4Q_{c}}{\pi^{2}r_{i}^{5}\sqrt{1+\frac{16Q_{c}^{2}}{b^{2}\pi^{4}r_{i}^{10}}}}\right)\left( \frac{2}{\Psi_{i}}\right)^{\frac{1}{2}}|Q-Q_{i}|^{\frac{1}{2}}.
\label{eq4.36}
\end{equation}

Comparing eqs.\eqref{eq4.25} and \eqref{eq4.36} we find that $ \delta=2 $.
\begin{flushleft}
\textit{\textbf{Critical exponent $ \phi $:}}
\end{flushleft}

The critical exponent $ \phi $ is associated with the divergence of the specific heat at constant charge ($ C_{Q} $) at the critical point $ r_{+}=r_{i} $ as,
\begin{equation}
C_{Q}\sim |Q-Q_{i}|^{-\phi}.
\label{eq4.37}
\end{equation}

Now from eqs.\eqref{eq4.7} and \eqref{eq4.11} we note that,
\begin{equation}
C_{Q}\sim \frac{1}{\Delta}
\label{eq4.38}
\end{equation}
which may be written as,
\begin{equation}
C_{Q}\sim \frac{1}{|Q-Q_{i}|^{1/2}}
\label{eq4.39}
\end{equation}
where we have used eq.\eqref{eq4.30}.

Comparison of eq.\eqref{eq4.39} with eq.\eqref{eq4.37} yields $ \phi=\dfrac{1}{2} $.

\begin{flushleft}
\textit{\textbf{Critical exponent $ \psi $:}}
\end{flushleft}

In order to calculate the critical exponent $ \psi $, which is related to the entropy of the third order LBI-AdS black hole, we Taylor expand the entropy ($ S(r_{+}) $) around the critical point $ r_{+}=r_{i} $. This gives,
\begin{equation}
S(r_{+})=S(r_{i})+\left[ \left( \frac{\partial S}{\partial r_{+}}\right) \right]_{r_{+}=r_{i}}(r_{+}-r_{i})+..... 
\label{eq4.40}
\end{equation}

If we neglect all the higher order terms and use eqs.\eqref{eq2.24}, \eqref{eq4.1} and \eqref{eq4.30}, we can write eq.\eqref{eq4.40} as,
\begin{equation}
S(r_{+})-S(r_{i})=\frac{5\pi^{3}}{4}\left(r_{i}^{4}+2\alpha' r_{i}^{2}+\alpha'^{2} \right) \left( \frac{2}{\Psi_{i}}\right)^{1/2}|Q-Q_{i}|^{1/2}. 
\label{eq4.41}
\end{equation}

Comparing eq.\eqref{eq4.41} with the standard relation 
\begin{equation}
S(r_{+})-S(r_{i})\sim |Q-Q_{c}|^{\psi}
\label{eq4.42}
\end{equation}
we finally obtain $ \psi=\dfrac{1}{2} $.

In the table below we write all the six critical exponents obtained from our analysis in a tabular form. For comparison we also give the critical exponents associated with some well known systems.\\
\begin{center}
\begin{tabular}{c c c c c c}
\hline Critical  & 3rd order & CrB$r_{3}$ $ ^{*}$  & 2D  & van der Waals's  & \\ 
exponents & LBI-AdS black hole & & Ising model$^{*}$ & system$ ^{\ddagger} $\\
\hline $\alpha$ & 0.5 & 0.05 & 0 & 0 & \\ 
 $\beta$ & 0.5 & 0.368 & 0.125 & 0.5 & \\ 
 $\gamma$ & 0.5 & 1.215 & 1.7 & 1.0 & \\ 
 $\delta$ & 2.0 & 4.28 & 15 & 3.0 & \\ 
 $\psi$ & 0.5 & 0.60 & - & - & \\ 
 $\phi$ & 0.5 & 0.03 & - & - & \\ 
\hline 
\end{tabular}
\end{center}
 
 ($ ^{*} $: these are the non-mean field values.)
 
 ($\ddagger$: these values are taken from \cite{Wu1}.)

\begin{flushleft}
\textbf{\textit{Thermodynamic scaling laws and static scaling hypothesis:}}
\end{flushleft}
The discussion of critical phenomena is far from complete unless we make a comment on the thermodynamic scaling laws. In standard thermodynamic systems the critical exponents are found to satisfy some relations among themselves. These relations are called thermodynamic scaling laws\cite{Hankey, Stanley}. These scaling relations are given as,
\begin{eqnarray}
\alpha + 2\beta +\gamma &=&2\nonumber\\
\alpha + \beta (\delta+1)&=&2\nonumber\\
\phi +2\psi - \frac{1}{\delta}&=&1\nonumber\\
\beta (\delta -1)&=&\gamma\nonumber\\
(2-\alpha)(\delta -1)&=&\gamma(1+\delta)\nonumber\\
1+(2-\alpha)(\delta\psi -1)&=&(1-\alpha)\delta
\label{eq4.43}
\end{eqnarray}

From the values of the critical exponents obtained in our analysis it is interesting to observe that all these scaling relations are indeed satisfied for the third order LBI-AdS black holes.

We shall now explore the static scaling hypothesis\cite{Hankey}-\cite{Lousto1} for the third order LBI-AdS black hole. Since we are working in the canonical ensemble framework, the thermodynamic potential of interest is the \textit{Helmhotz free energy}, $ \mathcal{F}(T,Q)=M-TS $, where the symbols have their usual meaning. 

Now the static scaling hypothesis states that,

 \textit{Close to the critical point the singular part of the Helmhotz free energy is a generalized homogeneous function of its variables.}

This asserts that there exist two parameters $ a_{\epsilon} $ and $ a_{\Upsilon} $ such that
\begin{equation}
\mathcal{F}(\lambda^{a_{\epsilon}}\epsilon , \lambda^{a_{\Upsilon}}\Upsilon)=\lambda \mathcal{F}(\epsilon,\Upsilon)
\label{eq4.44}
\end{equation}

for any arbitrary number $ \lambda $.

In an attempt to find the values of the scaling parameters $ a_{\epsilon} $ and $ a_{\Upsilon} $, we shall now Taylor expand the Helmhotz free energy $ \mathcal{F}(T,Q) $ near the critical point $ r_{+}=r_{i} $. This may be written as,
\begin{eqnarray}
\mathcal{F}(T,Q)&=&\mathcal{F}(T,Q)|_{r_{+}=r_{i}}+\left[ \left(\frac{\partial\mathcal{F}}{\partial T} \right)_{Q} \right]_{r_{+}=r_{i}}(T-T_{i})+\frac{1}{2}\left[ \left(\frac{\partial^{2}\mathcal{F}}{\partial T^{2}} \right)_{Q} \right]_{r_{+}=r_{i}}(T-T_{i})^{2}\nonumber\\
&&+\left[ \left(\frac{\partial\mathcal{F}}{\partial Q} \right)_{T} \right]_{r_{+}=r_{i}}(Q-Q_{i})+\frac{1}{2}\left[ \left(\frac{\partial^{2}\mathcal{F}}{\partial Q^{2}} \right)_{T} \right]_{r_{+}=r_{i}}(Q-Q_{i})^{2}\nonumber\\
&&+\left[\left(\frac{\partial^{2}\mathcal{F}}{\partial T\partial Q} \right) \right]_{r_{+}=r_{i}}(T-T_{i})(Q-Q_{i})+ .....
\label{eq4.45}
\end{eqnarray}

From eq.\eqref{eq4.45} we can identify the second derivatives of $ \mathcal{F} $ as,
\begin{equation}
\left(\frac{\partial^{2}\mathcal{F}}{\partial T^{2}} \right)_{Q}=\frac{-C_{Q}}{T}
\label{eq4.46}
\end{equation}

and \begin{equation}
\left(\frac{\partial^{2}\mathcal{F}}{\partial Q^{2}} \right)_{T}=\frac{\kappa_{T}^{-1}}{Q}.
\label{eq4.47}
\end{equation}

Note that, since both $ C_{Q} $ and $ \kappa_{T}^{-1} $ diverge at the critical point, these derivatives can be justified as the singular parts of the free energy $ \mathcal{F} $.

Since in the theory of critical phenomena we are mainly interested in the singular part of the relevant thermodynamic quantities, we sort out the singular part of $ \mathcal{F}(T,Q) $ from eq.\eqref{eq4.45}, which may be written as,
\begin{eqnarray}
\mathcal{F}_{s}&=&\frac{1}{2}\left[ \left(\frac{\partial^{2}\mathcal{F}}{\partial T^{2}} \right)_{Q} \right]_{r_{+}=r_{i}}(T-T_{i})^{2}+\frac{1}{2}\left[ \left(\frac{\partial^{2}\mathcal{F}}{\partial Q^{2}} \right)_{T} \right]_{r_{+}=r_{i}}(Q-Q_{i})^{2}\nonumber\\
&=&\frac{-C_{Q}}{2T_{i}}(T-T_{i})^{2}+\frac{\kappa_{T}^{-1}}{2Q_{i}}(Q-Q_{i})^{2}
\label{eq4.48}
\end{eqnarray}
where the subscript `$ s $' denotes the singular part of the free energy $ \mathcal{F} $.

Using eqs.\eqref{eq4.5}, \eqref{eq4.11}, \eqref{eq4.24} and \eqref{eq4.30} we may write the singular part of the Helmhotz free energy ($ \mathcal{F} $) as,
\begin{equation}
\mathcal{F}_{s}=\sigma_{i}\epsilon^{3/2}+\tau_{i}\Upsilon^{3/2}
\label{eq4.49}
\end{equation}

where 
\begin{equation}
\sigma_{i}=\frac{-\mathcal{A}_{i}T_{i}}{2}\,\,\,\,\,\,\, and,\,\,\,\,\,\,\,\tau_{i}=\frac{\mathcal{B}_{i}\Psi_{i}^{1/2}Q_{i}^{1/2}r_{i}}{2^{3/2}\Gamma_{i}^{1/2}}.
\label{eq4.50}
\end{equation}

From eqs.\eqref{eq4.44} and \eqref{eq4.49} we observe that
\begin{equation}
a_{\epsilon}=a_{\Upsilon}=\dfrac{2}{3}.
\label{eq4.51}
\end{equation}

This is an interesting result in the sense that, in general, $ a_{\epsilon} $ and $ a_{\Upsilon} $ are different for a generalized homogeneous function (GHF), but in this particular model of the black hole these two scaling parameters are indeed identical. With this result we can argue that the Helmhotz free energy is an usual homogeneous function for the third order LBI-AdS black hole. Moreover, we can determine the critical exponents ($ \alpha,\,\beta,\,\gamma,\,\delta,\,\phi,\,\psi $) once we calculate the scaling parameters. This is because these critical exponents are related to the scaling parameters as\cite{Hankey, Stanley},
\begin{eqnarray}
\alpha &=&2-\frac{1}{a_{\epsilon}},\qquad\beta = \frac{1-a_{\Upsilon}}{a_{\epsilon}},\nonumber\\
\gamma &=&\frac{2a_{\Upsilon}-1}{a_{\epsilon}},\qquad\delta =\frac{a_{\Upsilon}}{1a_{\Upsilon}},\nonumber\\
\phi &=&\frac{2a_{\epsilon}-1}{a_{\Upsilon}},\qquad\psi =\frac{1-a_{\epsilon}}{a_{\Upsilon}}.
\label{eq4.52}
\end{eqnarray}

There are two other critical exponents associated with the behavior of the \textit{correlation function} and \textit{correlation length} of the system near the critical surface. We shall denote these two critical exponents as $ \eta $ and $ \nu $ respectively. If $ G(\vec r_{+}) $ and $ \xi $ are the correlation function and the correlation length respectively, we can relate $ \eta $ and $ \nu $ with them as,
\begin{equation}
G({\vec r_{+}})\sim r_{+}^{2-n-\eta}
\label{eq4.53}
\end{equation}
and
\begin{equation}
\xi \sim |T-T_{i}|^{-\nu}.
\label{eq4.54}
\end{equation}

For the time being we shall assume that the two additional scaling relations\cite{Hankey}
\begin{equation}
\gamma = \nu(2-\eta)\,\,\,\,\,\,\, and \,\,\,\,\,\,\, (2-\alpha)=\nu n
\label{eq4.55}
\end{equation}
hold for the third order LBI-AdS black hole. Using these two relations (eq.\eqref{eq4.55}) and the values of $ \alpha $ and $ \gamma $, the exponents $ \nu $ and $ \eta $ are found to be $ \dfrac{1}{4} $ and $ 0 $ respectively.

Although we have calculated $ \eta $ and $ \nu $ assuming the additional scaling relations to be valid, it is not proven yet that these scaling relations are indeed valid for the black holes. One may adapt different techniques to calculate $ \eta $ and $ \nu $, but till now no considerable amount of progress has been made in this direction. One may compute these two exponents directly from the correlation of scalar modes in the theory of gravitation\cite{Jain} but the present theories of critical phenomena in black holes are far from complete.
\section{Conclusions}
In this paper we have analyzed the critical phenomena in higher curvature charged black holes in a canonical framework. For this purpose we have considered the third order Lovelock-Born-Infeld-AdS (LBI-AdS) black holes in a spherically symmetric space-time. We systematically derived the thermodynamic quantities for such black holes. We are able to show that some of the thermodynamic quantities ($ C_{Q}$, $ \kappa_{T}^{-1} $) diverge at the critical points. From the nature of the plots we argue that there is a higher order phase transition in this black hole. Although the analytical estimation of the critical points is not possible due to the complexity of the relevant equations, we are able to determine the critical points numerically. However, all the critical exponents are calculated analytically near the critical points.  Unlike the AdS black holes in the Einstein gravity, one interesting property of the higher curvature black holes is that the usual area law of entropy does not hold for these black holes. One might then expect that the critical exponents may differ form those for the AdS black holes in the Einstein gravity. But we find that all the critical exponents in the third order LBI-AdS black hole are indeed identical with those obtained in Einstein gravity\cite{DR2, DR1}. From this observation we may conclude that these black holes belong to the same universality class. Moreover, the critical exponents take the mean field values. It is to be noted that these black holes have distinct set of critical exponents which does not match with the critical exponents of any other known thermodynamic systems. Another point that must be stressed is that the static critical exponents are independent of the spatial dimensionality of the AdS space-time. This suggests the mean field behavior in black holes as thermodynamic systems and allows us to study the phase transition phenomena in the black holes. We have also discussed the static scaling laws and static scaling hypothesis. The static critical exponents are found to satisfy the static scaling laws near the critical points. We have checked the consistency of the static scaling hypothesis. Apart from this we note that the two scaling parameters have identical values. This allows us to conclude that the Helmhotz free energy is indeed a homogeneous function for this type of black hole. We have determined the two other critical exponents $ \nu $ and $ \eta $ associated with the correlation length ($ \xi $) and correlation function ($ G(\vec r_{+}) $) near the critical surface assuming the validity of the additional scaling laws. The values of these two exponents are found to be $ \dfrac{1}{4} $ and $ 0 $ respectively in the six spatial dimensions. Although the other six critical exponents are independent of the spatial dimension of the system, these two exponents are very much dimension dependent.

In our analysis we have been able to resolve a number of vexing issues concerning the critical phenomena in third order LBI-AdS black holes. But there still remains some unsolved problems that encourage one to make further investigations into the system. First of all, we have made a qualitative argument about the nature of the phase transition in this black hole. One needs to go through detailed algebraic analysis in order to determine the true order of the phase transition\cite{Modak1}-\cite{Lala}. Secondly, we have calculated the values of the exponents $ \nu $ and $ \eta $ assuming that the additional scaling relations hold for this black hole. But there is no evidence whether these two laws hold for the black hole\cite{DR2, DR1}. These scaling relations may or may not hold for the black hole. The dimension dependence of these two exponents ($ \nu $ and $ \eta $) makes the issue highly nontrivial in higher dimensions. A further attempt to determine $ \eta $ and $ \nu $ may be based on Ruppeiner's prescription \cite{Rup1}, where it is assumed that the absolute value of the thermodynamic scalar curvature ($ |\mathcal{R}| $) is proportional to the correlation volume $ \xi^{n} $:
\begin{equation}
|\mathcal{R}|\sim \xi^{n}
\label{eq4.56}
\end{equation}
where $n$ is the spatial dimension of the black hole. Now if we can calculate $ \mathcal{R} $ using the standard method\cite{Lala,Rup2, Rup3}, we can easily determine $ \xi $ from eq.\eqref{eq4.56}. Evaluating $ \xi $ around the critical point $ r_{+}=r_{i} $ as before, we can determine $ \nu $ directly. It is then straight forward to calculate $ \eta $ by using (80). This alternative approach, based on Ruppeiner's prescription, to determine $ \nu $ and $ \eta $ needs high mathematical rigor and also the complexity in the determination of the scalar curvature ($ \mathcal{R} $) in higher dimensions makes the issue even more challenging.

Apart from the above mentioned issues, it would also be highly nontrivial if we aim to investigate the AdS/CFT duality as an alternative approach to make further insight into the theory of critical phenomena in these black holes. The \textit{renormalization group} method may be another alternative way to describe the critical phenomena in these black holes.

Finally, it would be nice if we can apply the behaviour of the third order LBI-AdS black holes for understanding several issues related to the brane-world cosmology. Following the prescription of ref.\cite{BC_Rev}, we can embed a brane in the bulk third order LBI-AdS black hole and study the corresponding Friedmann-Robertson-Walker (FRW) cosmology. As a possible extension of our analysis, the study of thermodynamic properties of such a brane will be interesting. Performing analysis in the same line as is done in this paper we may compute the critical exponents of the brane which can be helpful to understand thermodynamical phases of the brane. On top of that, we can then compare critical behaviour of the black hole with that of the brane. Further, we may check the correspondence between the entropy of this black hole and that of the dual conformal field theory (CFT) that lives on the brane (Cardy-Verlinde formula) through the CFT/FRW relation. We may further look for possible modifications of the Cardy-Verlinde formula which may shed light on several questionable issues regarding our universe. Also, the contribution(s) due to the nonlinear Born-Infeld field, appearing in our model, on the FRW equations can be studied. There is also a scope to study the effects of the third order Lovelock coefficient ($\alpha'$) on the induced brane matter from the bulk LBI-AdS black hole. In relation with this, the investigation of the validity of the dominant energy condition (DEC) and/or weak energy condition (WEC) may turn out to be an important issue. 

\section*{Acknowledgements}
The author would like to thank the Council of Scientific and Industrial Research (C. S. I. R), Government of India, for financial support. He would also like to thank Prof. Rabin Banerjee, Dibakar Roychowdhury and Dr. Bibhas Ranjan Majhi for useful discussions.


\begin{thebibliography}{110}
\bibitem{Zumino}B. Zumino, Phys. Rep. 137, 109 (1986).
\bibitem{Green}M. B. Greens, J. H. Schwaz, E. Witten, Superstring Theory (Cambridge University Press).
\bibitem{Zwiebach}B. Zwiebach, Phys. Lett. 156B, 315 (1985).
\bibitem{Lovelock}D. Lovelock, J. Math. Phys. (N. Y.) 12, 498 (1971).
\bibitem{Deser}D. G. Boulware, S. Deser, Phys. Rev. Lett. 55, 2656 (1985).
\bibitem{seven6}H. L\"{u}, Y. Pang, Phys. Rev. D 81, 085016 (2010).
\bibitem{seven7}M. G\"{u}naydin, H. Nicolai, Phys. Lett. B 351, 169 (1995).
\bibitem{seven8}M. Rozali, Phys. Lett. B 400, 260 (1997).
\bibitem{seven9}Z-W. Chong, M. Cvetic, H. L\"{u}, C. N. Pope, Phys. Lett. B 626, 215 (2005).
\bibitem{seven10}D. D. K. Chow, Class. Quantum Grav. 25, 175010 (2008).
\bibitem{seven11}S-Q. Wu, Phys. Lett. B 705, 383 (2011).
\bibitem{Myres}R. C. Myres, M. J. Perry, Annals of Physics 172, 304-347 (1986).
\bibitem{Gregory}R. Gregory, R. Laflamme, Phys. Rev. Lett. 70, 2837 (1993).
\bibitem{Emparan}R. Emparan, H. S. Reall, Phys. Rev. Lett. 88, 101101 (2002).
\bibitem{Ban3}R. Banerjee, D. Roychowdhury, JHEP 11 (2011) 004.
\bibitem{DR2}R. Banerjee, D. Roychowdhury, Phys. Rev. D 85, 104043 (2012).
\bibitem{Emparan2}R. Emparan, H. S. Reall, Living Rev. Relativity 11, 6 (2008).
\bibitem{Bardeen}J. M. Bardeen, B. Carter, S. W. Hawking, Commun. Math. Phys. 31, 161-170 (1973).
\bibitem{Birrell}N. D. Birrell, P. C. W. Davies, Quantum Fields in Curved Space (Cambridge University Press, Cambridge, 1984).
\bibitem{Hawking-Page}S. W. Hawking, D. N. Page, Commun. Math. Phys. 87, 577 (1983).
\bibitem{Cham1}A. Chamblin, R. Emparan, C.V. Johnson, R.C. Myers, Phys. Rev. D 60, 064018 (1999).
\bibitem{Cai1}R. G. Cai, Phys. Rev. D 65, 084014 (2002).
\bibitem{Carlip}S. Carlip, S. Vaidya, Class. Quantum Grav. 20, 3827-3837 (2003).
\bibitem{Cai2}R. G. Cai, A. Wang, Phys. Rev. D 70, 064013 (2004).
\bibitem{Carter}B. M. N. Carter, I. P. Neupane, Phys. Rev. D 72, 043534 (2005) .
\bibitem{Cai3}R. G. Cai, S. P. Kim, B. Wang, Phys. Rev. D 76, 024011 (2007).
\bibitem{Myung2}Y. S. Myung, Mod. Phys. Lett. A 23, 667-676 (2008).
\bibitem{Modak1}R. Banerjee, S. K. Modak, S. Samanta, Eur. Phys. J. C 70: 317-328 (2010).
\bibitem{Modak2}R. Banerjee, S. K. Modak, S. Samanta, Phys. Rev. D 84, 064024 (2011).
\bibitem{Ban1}R. Banerjee, S. Ghosh, D. Roychowdhury, Phys. Lett. B 696, 156 (2011).
\bibitem{Ban2}R. Banerjee, S. K. Modak, D. Roychowdhury, JHEP 10 (2012) 125.
\bibitem{Lala}A.Lala, D. Roychowdhury, Phys. Rev. D 86, 084027 (2012).
\bibitem{Bibhas}B. R. Majhi, D. Roychowdhury, Class. Quantum Grav. 29, 245012 (2012).
\bibitem{Witten}E. Witten, Adv. Theor. Math. Phys. 2, 153 (1998).
\bibitem{Zeemansky}M. W. Zeemansky, R. H. Dittman, Heat and Thermodynamics: an intermediate textbook, McGraw-Hill.

\bibitem{Hankey}A. Hankey, H. E. Stanley, Phys. Rev. B 6, 3515 (1972).
\bibitem{Stanley}H. E. Stanley, Introduction to Phase Transitions and Critical Phenomena, Oxford University Press.
\bibitem{Lousto1}C. O. Lousto, Nucl. Phys. B 410, 155-172 (1993).
\bibitem{Lau}Y. K. Lau, Phys. Lett. A 186, 41 (1994).
\bibitem{Lousto2}C. O. Lousto, Gen. Relativ. Gravit. 27, 121 (1995).
\bibitem{Lousto3}C. O. Lousto, Phys. Rev. D 51, 1733 (1995).
\bibitem{Muniain}J. P. Muniain, D. Piriz, Phys. Rev. D 53, 816 (1996).
\bibitem{Cai4}R. G. Cai, Y. S. Myung, Nucl. Phys. B 495, 339-362 (1997).
\bibitem{Cai4a}R. G. Cai, Z-J. Lu, Y-Z. Zhang, Phys. Rev. D 55, 853 (1997).
\bibitem{Lousto4}C. O. Lousto, Int. J. Mod. Phys. D 6, 575-590 (1997).
\bibitem{Wu1}X. N. Wu, Phys. Rev. D 62, 124023 (2000).
\bibitem{Maeda}K. Maeda, M. Natssume, T. Okamura, Phys. Rev. D 78, 106007 (2008).
\bibitem{Jain}S. Jain, S. Mukherji, S. Mukhopadhyay, JHEP 0911:051 (2009).
\bibitem{Sahay1}A. Sahay, T. Sarkar, G. Sengupta, JHEP 1007:082 (2010).
\bibitem{Sahay2}A. Sahay, T. Sarkar, G. Sengupta, JHEP 1011:125 (2010).
\bibitem{Cai4b}Y. Liu, Q. Pan, B. Wang, R. G. Cai, Phys. Lett. B 693, 343350 (2010).
\bibitem{Wu2}C. Niu, Yu Tian, X. N. Wu, Phys. Rev. D 85, 024017 (2012).
\bibitem{DR1}R. Banerjee, D. Roychowdhury,  Phys. Rev. D 85, 044040 (2012).
\bibitem{Cvetic}M.~Cvetic, S.~Nojiri, S.~D.~Odintsov, Nucl.\ Phys.\ B 628, 295 (2002).
\bibitem{Cai5}R-G. Cai, Phys. Rev. D 65, 084014 (2002).
\bibitem{Banados}M. Banados, Phys. Lett. B 579, 13-24 (2004).
\bibitem{Dey}T. K. Dey, S. Mukherji, S. Mukhopadhyay, S. Sarkar, JHEP 04 (2007) 014.
\bibitem{Myung3}Y. S. Myung, Y-W. Kim, Y-J. Park, Eur. Phys. J. C 58: 337-346 (2008).
\bibitem{Anninos}D. Anninos, G. Pastras, JHEP 07 (2009) 030.
\bibitem{Cai6}Y. Liu, Q. Pan, B. Wang, R-G. Cai, Phys. Lett. B 693, 343-350 (2010).
\bibitem{Deh1}M. H. Dehghani, M. Shamirzaie, Phys. Rev. D 72, 124015 (2005).
\bibitem{Deh2}M. H. Dehghani, R. B. Mann,  Phys. Rev. D 73, 104003 (2006).
\bibitem{Aiello}M. Aiello, R. Ferraro, G. Giribet, Phys. Rev. D 70, 104014 (2004).
\bibitem{Deh3}M. H. Dehghani, N. Alinejadi, S. H. Hendi, Phys. Rev. D 77, 104025 (2008).
\bibitem{Carroll}S. Carroll, Spacetime and geometry, Addison Wesley, 2003.
\bibitem{Deh4}M. H. Dehghani, R. Pourhasan, Phys. Rev. D 79, 064015 (2009).
\bibitem{Cai7}R-G. Cai, Phys. Lett. B 582, 237-242 (2004).
\bibitem{Cai8}R-G. Cai, D-W. Pang, A. Wang, Phys. Rev. D 70, 124034 (2004).
\bibitem{Wald1}R. M. Wald, Phys. Rev. D 48, R3427 (1993).
\bibitem{Wald2}V. Iyer, R. M. Wald, Phys. Rev. D 50, 846 (1994).
\bibitem{Visser}M. Visser, Phys. Rev. D 48, 5697 (1993).
\bibitem{Jacobson1}T. Jacobson, R. C. Myres, Phys. Rev. Lett. 70, 3684 (1993).
\bibitem{Jacobson2}T. Jacobson, G. Kang, R. C. Myres, Phys. Rev. D 49, 6587 (1994).
\bibitem{Deh5}M. H. Dehghani, N. Bostani, A. Sheykhi, Phys. Rev. D 73, 104013 (2006).

\bibitem{Born}M. Born, L. Infeld, Proc. R. Soc. London A 144, 425 (1934).
\bibitem{Handbook}M. Abramowitz, I. A. Stegun, Handbook of Mathematical Functions, Dover, New York,
1972.
\bibitem{Malda1}J. Maldacena, Adv. Theor. Math. Phys. 2, 231 (1998).
\bibitem{Witt}E. Witten,  Adv. Theor. Math. Phys. 2, 253 (1998).
\bibitem{Malda2}O. Aharony, S. S. Gubser, J. Maldacena, H. Ooguri, Y. Oz, Phys. Rep. 323, 183 (2000).
\bibitem{Count1}M. H. Dehghani, R. B. Mann, Phys. Rev. D 64, 044003 (2001).
\bibitem{Count2}M. H. Dehghani, Phys. Rev. D 65, 104030 (2002).
\bibitem{Count3}A. Strominger, J. High Energy Phys. 10 (2001) 034.
\bibitem{Count4}A. Strominger, J. High Energy Phys. 11 (2001) 049.
\bibitem{Count5}V. Balasubramanian, P. Horova, D. Minic J. High Energy Phys. 05 (2001) 043.
\bibitem{Count6}S. Nojiri, S. D. Odintsov, Phys. Lett. B 519, 145 (2001).
\bibitem{Count7}S. Nojiri, S. D. Odintsov, J. High Energy Phys. 12(2001) 033.
\bibitem{Count8}R. G. Cai, Phys. Lett. B 525, 331 (2002).
\bibitem{Count9}R. Bousso, A. Maloney, A. Strominger Phys. Rev. D 65, 104039 (2002).
\bibitem{Count10}A. M. Ghezelbash, R. B. Mann, J. High Energy Phys. 01 (2002) 005.
\bibitem{Count11}M. H. Dehghani, Phys. Rev. D 65, 104003 (2002).
\bibitem{Brown}J. D. Brown, J. W. York, Phys. Rev. D 47, 1407 (1993).
\bibitem{Majhi_1}B. R. Majhi, T. Padmanabhan, Phys. Rev. D 85, 084040 (2012).
\bibitem{Majhi_2}B. R. Majhi, T. Padmanabhan, Phys. Rev. D 86, 101501(R) (2012).
\bibitem{B_Wang}S-J. Zhang, B. Wang, Phys. Rev. D 87, 044041 (2013).
\bibitem{Bekenstein}J. D. Bekenstein, Phys. Rev. D 7, 2333-2346 (1973).
\bibitem{Rup1}G. Ruppeiner, Phys. Rev. D 78, 024016 (2008).
\bibitem{Rup2}G. Ruppeiner, Phys. Rev. A 20, 1608 (1979).
\bibitem{Rup3}G. Ruppeiner, Rev. Mod. Phys. 67, 605 (1995); 68, 313(E) (1996).
\bibitem{BC_Rev}S.~Nojiri, S.~D.~Odintsov, S.~Ogushi, Int. Jour. Mod. Phys. A 32, 4809 (2002).
\end{thebibliography}
\end{document}